\documentclass[10pt,journal]{IEEEtran}

\usepackage{nomencl}
\makenomenclature

\renewcommand{\IEEEQED}{\IEEEQEDclosed}
\usepackage{color}

\usepackage{subfigure}
\usepackage{graphicx,cite,epsfig,amssymb,amsmath,multirow,lettrine,flushend,extarrows}
\usepackage{algorithm}
\usepackage{algorithmic}

\begin{document}
%
\title{Distributed Real-Time Energy Management in \\ Data Center Microgrids}

\author{{Liang~Yu,~\IEEEmembership{Member,~IEEE}, Tao~Jiang,~\IEEEmembership{Senior Member,~IEEE}, and Yulong Zou,~\IEEEmembership{Senior Member,~IEEE}}
\thanks{ \newline L. Yu and Y. Zou are with Key Laboratory of Broadband Wireless Communication and Sensor Network Technology of Ministry of Education, Nanjing University of Posts and Telecommunications, Nanjing 210003, P. R. China. (email: liang.yu@njupt.edu.cn) \newline
T. Jiang is Wuhan National Laboratory for Optoelectronics, School of Electronics Information and Communications, Huazhong University of Science and Technology, Wuhan 430074, P. R. China. \newline
}}


\maketitle

\begin{abstract}
Data center operators are typically faced with three significant problems when running their data centers, i.e., rising electricity bills, growing carbon footprints and unexpected power outages. To mitigate these issues, running data centers in microgrids is a good choice since microgrids can enhance the energy efficiency, sustainability and reliability of electrical services. Thus, in this paper, we investigate the problem of energy management for multiple data center microgrids. Specifically, we intend to minimize the long-term operational cost of data center microgrids by taking into account the uncertainties in electricity prices, renewable outputs and data center workloads. We first formulate a stochastic programming problem with the considerations of many factors, e.g., providing heterogeneous service delay guarantees for batch workloads, interactive workload allocation, batch workload shedding, electricity buying/selling, battery charging/discharging efficiency, and the ramping constraints of backup generators. Then, we design a realtime and distributed algorithm for the formulated problem based on Lyapunov optimization technique and a variant of alternating direction method of multipliers (ADMM). Moreover, the performance guarantees provided by the proposed algorithm are analyzed. Extensive simulation results indicate the effectiveness of the proposed algorithm in operational cost reduction for data center microgrids.
\end{abstract}

\IEEEpeerreviewmaketitle

\section{Introduction}\label{s1}

\IEEEPARstart{W}{ith} the development of Internet services and applications, massive geo-distributed data centers have been deployed. When running these data centers, a data-center operator is typically faced with three significant problems: (1) rising electricity bills, e.g., Google consumed 2260 GWh in 2010 and the corresponding electricity bill was larger than 1.35 billion dollars\cite{Gao2012}; (2) growing carbon emission, e.g., data center carbon emissions are expected to reach 2.6\% of the total emissions\cite{Gao2012}; (3) unexpected power outages, e.g., Amazon experienced several power outages during 2010-2013 and knocked many customers offline\cite{datacenterknowledge}. Since microgrids could potentially provide cost savings, emission reduction and reliability enhancement for data centers\cite{Salo2008,liangyu2015,LiangIEEEAccess2016,Li2016,Thompson2016}, it is necessary to study the problem of energy management for data center microgrids.

There has been few work on the energy management in microgrids. In \cite{Guan2010}, Guan \emph{et al.} investigated the scheduling problem of building energy supplies in a microgrid. In \cite{ErolKantarci2011}, Erol-Kantarci \emph{et al.} developed the idea of resource sharing among microgrids for the sake of increased reliability. In \cite{Huang2013}, Huang \emph{et al.} presented a novel energy management framework to minimize the operational cost of a microgrid by introducing a model of QoSE (quality-of-service in electricity). In \cite{Zhangyu2013}, Zhang \emph{et al.} considered an optimal energy management problem for both supply and demand of a grid-connected microgrid incorporating renewable energy sources. In \cite{ZhangRui2015}, Zhang \emph{et al.} proposed an online algorithm to minimize the total energy cost of the conventional energy drawn from the main grid over a finite horizon by scheduling energy storage devices in a microgrid. In \cite{Ma2016}, Wang \emph{et al.} designed a distributed algorithm for online energy management in networked microgrids with a high penetration
of distributed energy resources using online ADMM with Regret. In \cite{GuoTSG2016}, Guo \emph{et al.} proposed a two-stage adaptive robust optimization approach for the energy management of a microgrid. In \cite{Salo2008}, Salomonsson \emph{et al.} designed an adaptive control system for a dc microgrid with data center loads. In \cite{Shi2016}, Shi \emph{et al.} proposed an online energy management strategy for realtime operation of a microgrid with the considerations of the power flow and system operational constraints on a distribution network. In \cite{Li2016}, Li \emph{et al.} studied the problem of minimizing the operation cost of a data center microgrid. In \cite{Chen2016}, Chen \emph{et al.} proposed a cooling-aware realtime algorithm to minimize the long-term operational cost of a data center microgrid. In \cite{Thompson2016}, Thompson \emph{et al.} presented a methodology for optimizing investment in data center battery storage capacity in a microgrid. Though some positive results have been obtained in the above works, there is no work that focuses on the realtime and distributed energy management for multiple data center microgrids. In our previous works\cite{liangyu2015}\cite{LiangIEEEAccess2016}\cite{Yu2014iot}, we mainly focus on the realtime energy management for multiple data center microgrids from different perspectives, e.g., energy cost reduction and carbon emission reduction. However, such previous works neglect heterogeneous service delay guarantees for batch workloads in all data centers\cite{LiangTPDSdelay2015} and distributed implementation for the proposed realtime algorithm.

Based on the above observation, this paper investigates the problem of realtime distributed energy management for multiple data center microgrids considering the drawbacks in our previous works. The resulting challenge consists of two aspects, i.e., spatial and temporal couplings\cite{dengR2015}. On one hand, there are some spatial couplings among all microgrids due to the allocation of interactive workloads. On the other hand, to provide the heterogeneous service delay guarantees for batch workloads in all data centers and keep all energy storage systems stable, several temporal couplings are incurred.

To deal with the above challenge, we first formulate a stochastic programming problem to minimize the time average expected operational cost by jointly capturing the constraints with geographical load balancing, batch workload allocation/shedding, heterogeneous service delay guarantees for batch workloads, electricity buying/selling, battery charging/discharging management, backup generators, and power balancing. Since the formulated optimization problem is a large-scale nonlinear stochastic programming with ``time-coupling" constraints, we propose a realtime and distributed algorithm based on Lyapunov optimization technique\cite{Neely2010} and a variant of alternating direction method of multipliers (ADMM)\cite{He2012}\footnote{In \cite{Sunsun2014}, Sun \emph{et al.} adopted Lyapunov optimization technique and ADMM with two blocks to deal with the problem of power balancing in a renewable-integrated power grid with storage and flexible loads. Though data centers could be also regarded as a kind of flexible loads, it does not mean that the method in \cite{Sunsun2014} could be applied to our problem directly. The reason is that this paper considers both multiple data center microgrids and heterogeneous service delay guarantees for batch workloads in all data centers. Specifically, to provide the heterogeneous service delay guarantees for batch workloads in all data centers and keep all energy storage systems stable, we adopt three queues and define a weighted quadratic Lyapunov function when proposing a realtime algorithm with Lyapunov optimization technique. Moreover, we design a distributed implementation of the proposed realtime algorithm based on ADMM with eleven blocks.}. The key idea of the proposed algorithm is given as follows. Firstly, we propose a realtime algorithm for the formulated problem based on Lyapunov optimization technique so that ``time-coupling" constraints could be avoided. Then, we present the distributed implementation of the proposed realtime algorithm without considering the nonlinear constraints based on a variant of ADMM. Next, a feasible solution to the original problem could be obtained by adjustment so that the nonlinear constraints in the formulated problem could be satisfied. Furthermore, the performance analysis of the proposed algorithm is carried out.

The main contributions of this paper are summarized below:
\begin{itemize}
  \item We formulate a stochastic programming to minimize the long-term operational cost of multiple data center microgrids with the considerations of many factors, e.g., providing heterogeneous service delay guarantees for batch workloads, interactive workload allocation, batch workload shedding, electricity buying/selling, battery charging/discharging efficiency, and the ramping constraints of backup generators.
  \item We propose a realtime and distributed algorithm to solve the formulated problem based on Lyapunov optimization technique and a variant of ADMM. Moreover, we analyze the performance guarantees provided by the proposed algorithm. Note that the proposed algorithm does not require any prior knowledge of statistical characteristics associated with system parameters and has low computational complexity.
  \item We conduct extensive simulations to evaluate the performance of the proposed algorithm. Simulation results show that the proposed algorithm outperforms other benchmark schemes in operational cost reduction.
\end{itemize}

The rest of this paper is organized as follows. In Section \ref{s2}, we describe the system model and problem formulation. Section \ref{s3} proposes a realtime and distributed algorithm to solve the formulated problem. Section \ref{s4} gives the algorithmic performance analysis. Extensive simulations are conducted in Section \ref{s5}. Finally, conclusions are drawn in Section \ref{s6}.

\section{Model And Formulation}\label{s2}
\begin{figure}[!htb]
\centering
\includegraphics[scale=0.6]{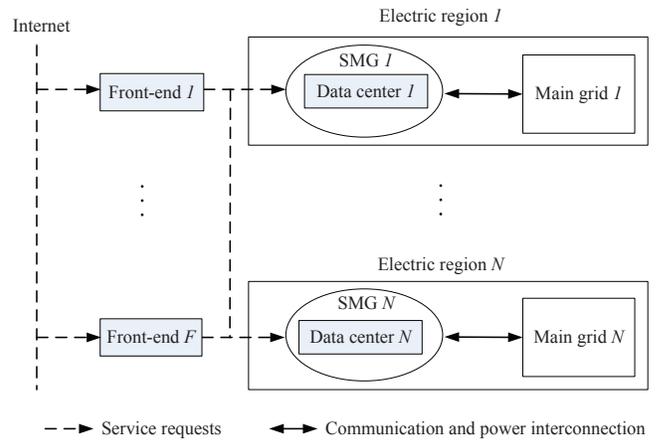}
\caption{System model}\label{fig_1}
\end{figure}

We consider a data center operator that has some geo-distributed data centers located in different electric regions as shown in Fig.~\ref{fig_1}, where each data center operates in a smart microgrid (SMG) environment\cite{ErolKantarci2011}. As far as the operation condition of a SMG is concerned, there are two modes, i.e., the islanded mode and the grid-connected mode. In the islanded mode, SMGs could supply their loads using multiple energy resources, e.g., energy storage devices, renewable and backup generators. In contrast, a SMG could sell (buy) energy to (from) a main grid in the grid-connected mode. A SMG considered in this paper consists of four main components, i.e., a generation system, a load, an energy storage system (ESS), and an energy management system (EMS). Specifically, a generation system consists of several renewable generators and a conventional generator (usually adopted as the backup generator), while the EMS is responsible for the energy scheduling of other components in the SMG. As the aggregated load in the SMG, a data center needs to finish the interactive workloads dispatched from front-end servers and the batch workloads within the data center. In this paper, we consider a time-slotted system and the length of each slot is assumed to be unit time. For easy reading, the main notations are introduced in Table \ref{table_1}.

\begin{table}[!t]
\renewcommand{\arraystretch}{1.3}
\scriptsize
\caption{Notations}
\label{table_1}
\centering
\begin{tabular}{|c|c|}
\hline
 Notation &  Definition\\
\hline\hline
t & Time slot index ($1\leq t \leq T$)\\
\hline
f & front-end server index ($1\leq f \leq F$) \\
\hline
$i$ & A common index for data centers, SMGs and main grids\\
\hline
$f$ & Front-end server $f$\\
\hline
$\lambda_{f,t}$ & The number of interactive workloads at front-end server $f$ at $t$\\
\hline
$d_{f,i,t}$ & Interactive workload allocation from front-end server $f$ to DC $i$ at $t$\\
\hline
$\pi_{i,q,t}$ &  The quantity of batch workloads with type $q$ at $t$ ($1\leq q\leq M_i$)\\
\hline
$Q_{i,q,t}$ & Batch workload queue \\
\hline
$x_{i,q,t}$ & The served workloads in batch workload queue at $t$ \\
\hline
$e_{i,q,t}$ & The quantity of dropped batch workloads at $t$ \\
\hline
$R_{i,q}^{\max}$ & The maximum queueing delay associated with $\pi_{i,q,t}$  \\
\hline
$\mathcal{T}_{i,q}$ & The tolerant service delay associated with $\pi_{i,q,t}$  \\
\hline
$P_{i,\text{idle}}$ & Idle power of a server in data center $i$ \\
\hline
$P_{i,\text{peak}}$ & Peak power of a server in data center $i$ \\
\hline
$p_{i,t}$ & Total power consumption in data center $i$ at $t$ \\
\hline
$r_{i,t}$ &  The total power output of the renewable generators in SMG $i$ at $t$\\
\hline
$c_{i,t}$ &  The power output of the conventional generator in SMG $i$ at $t$\\
\hline
$\epsilon_i$ & Ramping coefficient of the conventional generator in SMG $i$ \\
\hline
$u_{c,i,t}$ & The charging power for the ESS in SMG $i$ at $t$\\
\hline
$u_{d,i,t}$ & The discharging power for the ESS in SMG $i$ at $t$\\
\hline
$D_{i,t}$ & The stored energy level of the ESS $i$ at $t$\\
\hline
$S_i(t)$ & Purchasing electricity price from main grid $i$ at $t$ \\
\hline
$W_i(t)$ & Selling electricity price to main grid $i$ at $t$ \\
\hline
$g_{i,t}$ & Energy transactions between SMG $i$ and main grid $i$ at $t$ \\
\hline
$\Gamma_{1,t}$ & The cost incurred by electricity buying and selling at $t$ \\
\hline
$\Gamma_{2,t}$ & Total revenue loss of serving interactive requests at $t$ \\
\hline
$\Gamma_{3,t}$ & The penalty cost imposed on dropping batch workloads at $t$ \\
\hline
$\Gamma_{4,t}$ & Battery depreciation cost at $t$ \\
\hline
$\Gamma_{5,t}$ & Generation cost of the conventional generators at $t$ \\
\hline
$H_{i,q,t}$ & Delay-aware virtual queue \\
\hline
$Z_{i,t}$ & Virtual energy queue \\
\hline
$\Delta_t$ & one-slot conditional Lyapunov drift \\
\hline
$\Delta V_t$ & drift-plus-penalty term \\
\hline
\end{tabular}
\end{table}

\subsection{Models Associated with Data Centers and Front-end Servers}
Suppose that there are $N$ data centers geographically distributed in $N$ SMGs, which connected to $N$ main grids. Therefore, a common index $i$ ($1\leq i\leq N$) is adopted for data centers, SMGs and main grids. Moreover, we assume that data center $i$ consists of $C_i$ homogeneous servers\footnote{Although all the servers at a data center are assumed to be homogeneous, the model could be extended to the case with heterogeneous servers by adopting a few additional notations.}. In time slot $t$, the total quantity of interactive workloads (in the number of servers required) at the front-end server $f$ ($1\leq f\leq F$) is $\lambda_{f,t}$. Let $d_{f,i,t}$ be the quantity of interactive workloads allocated from front-end server $f$ to data center $i$ at slot $t$. Then, we have\cite{LiangTCC2016}\cite{LiangRisk2014}
\begin{align} \label{f_1}
&\sum\limits_{i=1}^N d_{f,i,t}=\lambda_{f,t},~\forall f,t, \\
&~~~d_{f,i,t}\geq 0,~\forall f,i,t.
\end{align}

Besides interactive workloads, some resource elastic batch workloads are commonly processed within data centers, e.g., scientific applications, data mining jobs. Batch workloads could be scheduled at any time slot as long as they are processed before their deadlines. Thus, batch workloads could be buffered and served in proper time slot. Let $\pi_{i,q,t}$ be the quantity of batch workloads at slot $t$ (also in terms of the number of servers required) with type $q$ ($1\leq q\leq M_i$) in data center $i$. By storing batch workloads $\pi_{i,q,t}$ in a queue $Q_{i,q,t}$ according to its type $q$, we have
\begin{align} \label{f_2}
Q_{i,q,t+1}=\max[Q_{i,q,t}-x_{i,q,t},0]+\pi_{i,q,t},~\forall i,q,t
\end{align}
where $x_{i,q,t}$ denotes the served workloads in the queue $q$ of data center $i$ at slot $t$. Denote the maximum value of $x_{i,q,t}$ by $x_{i,q}^{\max}$, where $x_{i,q}^{\max}\geq \pi_{i,q}^{\max}$ ($\pi_{i,q}^{\max}=\max_t \pi_{i,q,t}$) so that it is always possible to make the queue $Q_{i,q,t}$ stable (and this can be done with one slot delay if we choose $x_{i,q,t}=x_{i,q}^{\max}$ for all $t$). In addition, by observing the structure of $Q_{i,q,t}$, it can be found that there is no need to serve the batch workload that is larger than $Q_{i,q,t}$. Thus, we have
\begin{align}\label{f_3}
0\leq x_{i,q,t}\leq \min\{x_{i,q}^{\max},Q_{i,q,t}\},~\forall i,q,t.
\end{align}

To keep workload queues $Q_{i,q,t}$ stable, the batch workloads should be served without waiting for a long time. Since the summation of served batch workloads $\sum\nolimits_{q=1}^{M_i}x_{i,q,t}$ and arrived interactive workloads $\sum\nolimits_{f=1}^F d_{f,i,t}$ may exceed the processing capacity of data center $i$, some batch workloads have to be dropped at this time. Let $e_{i,q,t}$ be the quantity of dropped batch workloads, we have
\begin{align} \label{f_4}
&\sum\limits_{f=1}^F d_{f,i,t}+\sum\limits_{q=1}^{M_i}(x_{i,q,t}-e_{i,q,t}) \leq C_i,~\forall i,t.\\
&~~~~~~~~~~~0\leq e_{i,q,t}\leq x_{i,q,t},~\forall i,q,t.
\end{align}

For any control algorithm, it is necessary to ensure that the average length of the workload queue $q$ in data center $i$ is finite so that batch workloads could be finished without waiting an arbitrarily long time, i.e.,
\begin{align}\label{f_5}
\overline{Q}_{i,q}=\mathop {\lim\sup}\limits_{T \to \infty}\frac{1}{T}\sum\nolimits_{t=0}^{T-1} \mathbb{E}\{Q_{i,q,t}\}<\infty.
\end{align}

Note that \eqref{f_5} is not enough to ensure the heterogeneous service delay for batch workload $\pi_{i,q,t}$, we adopt the following constraint,
\begin{align} \label{f_6}
R_{i,q}^{\max}\leq \mathcal{T}_{i,q},~\forall i,q,t
\end{align}
where $R_{i,q}^{\max}$ and $\mathcal{T}_{i,q}$ are the maximum queueing delay and the tolerant service delay associated with the batch workload added into the queue $Q_{i,q,t}$ of data center $i$ at slot $t$, respectively. In Section \ref{s5}, we will provide the specific expression of $R_{i,q}^{\max}$.

Let $\text{PUE}_i$ be the $\text{PUE}$\footnote{PUE is defined as the ratio of the total power consumption at a data center to the power consumption at IT equipments} of data center $i$, $P_{i,\text{idle}}$ and $P_{i,\text{peak}}$ represent the idle power and peak power of a server in data center $i$, respectively. Then, the total power consumption in data center $i$ at slot $t$ $p_{i,t}$ could be estimated by\cite{Qureshi2009}
\begin{align}\label{f_7}
p_{i,t}=\alpha_i+ \beta_i\Big(\sum\limits_{f=1}^F d_{f,i,t}+\sum\limits_{q=1}^{M_i}(x_{i,q,t}-e_{i,q,t})\Big),~\forall i,t,
 \end{align}
where $\alpha_i  \triangleq C_i(P_{i,\text{idle}} + (\text{PUE}_i - 1)P_{i,\text{peak}})$, $\beta_i \triangleq P_{i,\text{peak}} - P_{i,\text{idle}}$.

\subsection{Models Related to the Generation System and ESS}
\subsubsection{Generation Model}
Let $r_{i,t}$ and $c_{i,t}$ be the total power output of the renewable generators and the power output of the conventional generator in SMG $i$ at slot $t$, respectively. Then, we have
\begin{align}\label{f_8}
0\leq c_{i,t} \leq c_{i,\max},~\forall i,t,
 \end{align}
where $c_{i,\max}$ is the maximum power output associated with the conventional generator in SMG $i$. Considering the physical constraints of the conventional generator, the output change in two consecutive slots is limited instead of arbitrarily large, which is reflected by a so-called ramping constraint. Without loss of generality, the ramp-up and ramp-down constraints are regarded as the same\cite{Sunsun2014}. Then, we have
\begin{align}\label{f_9}
|c_{i,t}-c_{i,t-1}| \leq \epsilon_i c_{i,\max},~\forall i,t,
 \end{align}
where $\epsilon_i$ is the ramping coefficient associated with the conventional generator in SMG $i$.

\subsubsection{ESS Model}
We define $u_{c,i,t}$ and $u_{d,i,t}$ to represent the charging and discharging power for the ESS in SMG $i$ at slot $t$. Then, we have
\begin{align}\label{f_10}
0\leq u_{c,i,t}\leq u_{i,\text{cmax}},~\forall i,t,\\
0\leq u_{d,i,t}\leq u_{i,\text{dmax}},~\forall i,t,
\end{align}
where $u_{i,\text{cmax}}$ and $u_{i,\text{dmax}}$ are maximum charging power and discharging power, respectively. Denote $\eta_{c,i}$ and $\eta_{d,i}$ be the charging and discharging efficiency of the ESS in SMG $i$ at slot $t$, respectively. In addition, simultaneous charging and discharging are not allowed considering the round-trip inefficiency, i.e.,
\begin{align}\label{f_11}
u_{c,i,t}\cdot u_{d,i,t}=0,\forall i,t.
\end{align}

Let $D_{i,t}$ be the stored energy of the ESS $i$, we have
\begin{align}\label{f_12}
D_{i,\min}\leq D_{i,t}\leq D_{i,\max},~\forall i,t,
\end{align}
where $D_{i,\max}$ and $D_{i,\min}$ denote the maximum and the minimum capacity of the ESS $i$, respectively. In addition, the storage dynamics of the ESS $i$ could be modeled by
\begin{align}\label{f_13}
D_{i,t+1}=D_{i,t}+ \eta_{c,i} u_{c,i,t}-\frac{1}{\eta_{d,i}}u_{d,i,t},~\forall i,t.
\end{align}

To satisfy the energy demand of data centers, SMGs may exchange energy with main grids. Denote the electricity price of buying and selling energy by $X_{i,t}\in [X_{i,\min},~X_{i,\max}]$ and $W_{i,t} \in [W_{i,\min},~W_{i,\max}]$, respectively. As in \cite{Zhangyu2013}, the selling price is assumed to be strictly smaller than the purchasing price so that energy arbitrage could be avoided, i.e., $X_{i,t}>W_{i,t}$. To achieve the real-time power balancing, we have the following constraints, i.e.,
\begin{align}\label{f_14}
g_{i,t}+r_{i,t}+c_{i,t}+u_{d,i,t}=p_{i,t}+u_{c,i,t},~\forall i,t,
\end{align}
where $g_{i,t}$ denotes the energy transactions between SMG $i$ and main grid $i$ at slot $t$, which is bounded by
\begin{align}\label{f_15}
G_{i,\text{smax}}\leq g_{i,t}\leq G_{i,\text{bmax}},~\forall i,t,
\end{align}
where $G_{i,\text{bmax}}>0$ and $G_{i,\text{smax}}<0$ are determined by the physical limitations, e.g., transmission lines\cite{Huang2013}. As in \cite{liangyu2015}, $G_{i,\text{bmax}}$ and $G_{i,\text{smax}}$ are assumed to be large enough to support the normal operation of SMG $i$ in the grid-connected mode.

\subsection{Operational Cost Model}\label{s37}
Denote the total operational cost of the data center operator at slot $t$ by $\Gamma_t$, which includes several components, i.e., the cost of purchasing and selling electricity $\Gamma_{1,t}$, revenue loss associated with workload allocation $\Gamma_{2,t}$ and $\Gamma_{3,t}$, the battery depreciation cost $\Gamma_{4,t}$, and the total generation cost of conventional generators $\Gamma_{5,t}$. Specifically, the cost incurred by electricity buying and selling at slot $t$ $\Gamma_{1,t}$ is obtained below,
\begin{align}\label{f_16}
&\Gamma_{1,t}=\sum\limits_{i=1}^{N}\Bigg(\frac{X_{i,t}-W_{i,t}}{2}|g_{i,t}|+\frac{X_{i,t}+W_{i,t}}{2}g_{i,t}\Bigg).
\end{align}

For interactive applications, latency is the most important performance metric and a moderate increase in user-perceived
latency would translate into substantial revenue loss for the data center operator\cite{Xu2013}\cite{Zhou2015infocom}. To model the utility of the interactive workload, the convex function in \cite{Xu2013} is adopted, which converts the mean propagation delay into revenue loss, i.e., $\omega\sum\nolimits_{i=1}^N d_{f,i,t}L_{f,i}/\lambda_{f,t}$, where $\omega$ is a conversion factor; $L_{f,i}$ is the propagation latency between the front-end server $f$ and data center $i$. Then, the total revenue loss of serving interactive requests is described by $\Gamma_{2,t}=\omega\sum\nolimits_{f=1}^F\sum\nolimits_{i=1}^N d_{f,i,t}L_{f,i}$.

In addition, to model the revenue loss of allocating processing servers for batch workload, the following function is adopted as in\cite{Zhou2015infocom},
$\Gamma_{3,t}=\sum\nolimits_{i=1}^N\sum\nolimits_{q=1}^{M_i}\theta_i e_{i,q,t}$, where $\theta_i$ is the penalty factor imposed on dropping batch workloads.

It is known that charging and discharging of batteries would affect their lifetime. To model such depreciation cost, the penalty function $B_i(u_{c,i,t},u_{d,i,t})$ is adopted. Continually, we have $\Gamma_{4,t}=\sum\nolimits_{i=1}^N B_i(u_{c,i,t},u_{d,i,t})$.

Denote the generation cost function of the conventional generator at slot $t$ by $A_i(c_{i,t})$. Then, $\Gamma_{5,t}=\sum\nolimits_{i=1}^N A_i(c_{i,t})$.

With the above-mentioned cost components, the total operational cost of the data center operator is calculated by $\Gamma_t=\sum\nolimits_{l=1}^5 \Gamma_{l,t}$.

\subsection{Operational Cost Minimization Problem}
With the aforementioned models, we can formulate a stochastic programming problem to minimize the time average expected operational cost of data center microgrids as follows,
\begin{subequations}\label{f_17}
\begin{align}
(\textbf{P1})~~&\min~\mathop {\lim\sup}\limits_{T \to \infty}\frac{1}{T}\sum\limits_{t=0}^{T-1} \mathbb{E}\{\Gamma_t\},  \\
s.t.&~\eqref{f_1}-\eqref{f_15},
\end{align}
\end{subequations}
where $\mathbb{E}\{\cdot\}$ is the expectation operator; the decision variables are $d_{f,i,t}$, $x_{i,q,t}$, $e_{i,q,t}$, $c_{i,t}$, $g_{i,t}$, $u_{c,i,t}$ and $u_{d,i,t}$; the expectation in the objective function is taken over the randomness of the system parameters $\lambda_{f,t}$, $\pi_{i,q,t}$, $r_{i,t}$, $X_{i,t}$ and $W_{i,t}$, and the possibly random control actions at each time slot.

For simplicity, the cost functions $A_i(\cdot)$ and $B_i(\cdot)$ are assumed to be continuously differentiable and convex, which is reasonable since many practical costs could be well approximated by such functions\cite{ZhangRui2015}\cite{Rivera2013}. Let $A_i'(\cdot)$ and $B_i'(\cdot)$ be the derivatives of $A_i(\cdot)$ and $B_i(\cdot)$, respectively. In addition, we suppose that $A_i'(c_{i,t})$ and $B_i'(u_{c,i,t},u_{d,i,t})$ are bounded within the intervals [$A_{i,\min}'$,~$A_{i,\max}'$] and [$B_{i,\min}'$,~$B_{i,\max}'$], respectively.

\begin{figure*}[!htb]
\centering
\includegraphics[scale=1]{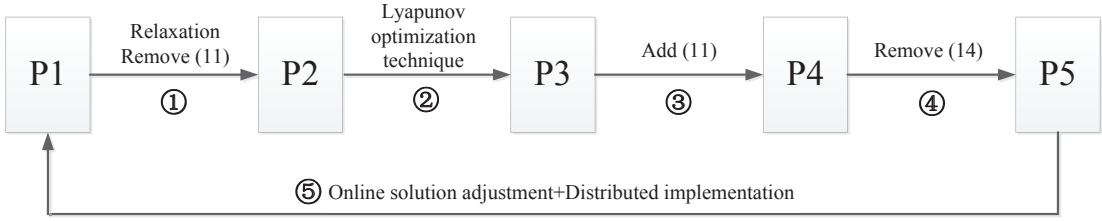}
\caption{An illustration of the key idea of the proposed algorithm}\label{fig_2}
\end{figure*}

\section{Algorithm Design}\label{s3}
There are three challenges to solve \textbf{P1}. Firstly, \textbf{P1} is a large-scale nonlinear optimization problem as the data center operator may deploy tens of geo-distributed data centers and hundreds of thousands of front-end servers around the world. Secondly, the future parameters are not known, including workload, renewable generation output and electricity price. Thirdly, the constraints \eqref{f_9} and \eqref{f_13} bring the ``time coupling" property to \textbf{P1}, which means that the current decision can impact the future decision. Previous methods to handle the ``time coupling" problem are usually based on dynamic programming, which suffers from ¡°the curse of dimensionality¡± problem. The structure and size of \textbf{P1} motivates us to design a scalable distributed realtime algorithm that is applicable for practical applications.

The key idea of the proposed algorithm can be illustrated by Fig.~\ref{fig_2}. Specifically, we can first transform the original problem \textbf{P1} into a stochastic programming problem \textbf{P2} with time average constraints by removing the constraint \eqref{f_9}. Then, we can transform \textbf{P2} into one-slot minimization problem \textbf{P3} using Lyapunov optimization technique. Next, by incorporating the constraint \eqref{f_9} into \textbf{P3}, we obtain \textbf{P4}. Since there are nonlinear constraints \eqref{f_11} in \textbf{P4}, we transform \textbf{P4} into \textbf{P5} by removing \eqref{f_11}. After obtaining the solution of \textbf{P5}, we adjust the solution so that \eqref{f_11} could be satisfied. Finally, we provide the distributed implementation of the proposed online algorithm and prove that all constraints of \textbf{P1} could be satisfied by the proposed algorithm.

Since Lyapunov optimization technique (LOT) could be used to solve a stochastic programming problem with time average constraints, we need to transform (15) and (16) into the time average constraints. To be specific, we define $\overline{u_{c,i}}$ and $\overline{u_{d,i}}$ as follows,
\begin{align}\label{f_18}
\overline {u_{c,i}}  = \mathop {\lim \sup }\limits_{T \to \infty } \frac{1}{T}\sum\nolimits_{t = 0}^{T - 1} \mathbb{E}\{u_{c,i,t}\}, \\
\overline {u_{d,i}}  = \mathop {\lim \sup }\limits_{T \to \infty } \frac{1}{T}\sum\nolimits_{t = 0}^{T - 1} \mathbb{E}\{u_{d,i,t}\}.
\end{align}

It is not difficult to obtain that $\eta_{c,i}\overline{u_{c,i}}=\frac{1}{\eta_{d,i}}\overline{u_{d,i}}$. Continually, \textbf{P1} could be relaxed into \textbf{P2} below,
\begin{subequations}\label{f_23}
\begin{align}
(\textbf{P2})~~&\min~\mathop {\lim\sup}\limits_{T \to \infty}\frac{1}{T}\sum\limits_{t=0}^{T-1} \mathbb{E}\{\Gamma_t\},  \\
s.t.&~(1),(2),(4)-(10),(12)-(14),(17),(18), \\
&\eta_{c,i}\overline{u_{c,i}}=\frac{1}{\eta_{d,i}}\overline{u_{d,i}},~\forall i.
\end{align}
\end{subequations}

To solve \textbf{P2}, LOT intends to transform time average constraints into queue stability problems. Thus, a virtual energy queue $Z_{i,t}$ is adopted to ensure the feasibility of $\eta_{c,i}\overline{u_{c,i}}=\frac{1}{\eta_{d,i}}\overline{u_{d,i}}$, i.e.,
\begin{align}\label{f_19}
Z_{i,t}=D_{i,t}-D_{i,\min}-V\eta_{d,i}\gamma_{i,\max}-\frac{1}{\eta_{d,i}}u_{i,\text{dmax}},
\end{align}
where $\gamma_{i,\max}$=$\max\{X_{i,\max},W_{i,\max},A_{i,\max}'\}$; $V\in [0,V_{\max}]$ is a control parameter that would be specified later. Continually, the update equation of $Z_{i,t}$ is obtained as follows,
\begin{align}\label{f_20}
Z_{i,t+1}=Z_{i,t}+\eta_{c,i} u_{c,i,t}-\frac{1}{\eta_{d,i}}u_{d,i,t},~\forall i,t.
\end{align}

Similarly, to ensure the feasibility of (7), we need to keep the workload queue $Q_{i,q,t}$ stable. In addition, to ensure the feasibility of \eqref{f_6}, we adopt a delay-aware virtual queue $H_{i,q,t}$. Specifically, for each $i$ and $q$, $H_{i,q,t}$ with $H_{i,q,0}=0$ and with dynamics as follows,
\begin{align} \label{f_21}
H_{i,q,t + 1}  = \left\{ \begin{array}{l}
 [H_{i,q,t}  - x_{i,q,t}  + \varepsilon_{i,q}]^ +, Q_{i,q,t}> x_{i,q,t}, \\
 0,\begin{array}{*{20}c}
   {}  \\
\end{array}\begin{array}{*{20}c}
   {}  \\
\end{array}\begin{array}{*{20}c}
   {}  \\
\end{array}\begin{array}{*{20}c}
   {}  \\
\end{array}\begin{array}{*{20}c}
   {}  \\
\end{array}\begin{array}{*{20}c}
   {}  \\
\end{array}\begin{array}{*{20}c}
   {}  \\
\end{array}\begin{array}{*{20}c}
   {}  \\
\end{array}\begin{array}{*{20}c}
   {}  \\
\end{array}Q_{i,q,t}  \le x_{i,q,t}, \\
 \end{array} \right.
\end{align}
where $[\diamond]^+\triangleq \max\{\diamond,0\}$; $\varepsilon_{i,q}$ is a fixed parameter, which would be specified later. It can be observed that $H_{i,q,t+1}$ has the same service rate as $Q_{i,q,t+1}$ but has a new arrival rate $\varepsilon_{i,q}$ when $Q_{i,q,t}> x_{i,q,t}$, which can ensure that $H_{i,q,t+1}$ grows when the batch workload added into the queue $Q_{i,q,t}$ at slot $t$ is still waiting to be satisfied. If we can ensure that the queues $H_{i,q,t}$ and $Q_{i,q,t}$ have finite upper bounds, then the maximum queueing delay in queue $Q_{i,q,t}$ defined in the following Lemma could be guaranteed.

\textbf{\emph{Lemma 1~(Maximum Queueing Delay)}}
Suppose we can control the system so that $H_{i,q,t}\leq H_{i,q}^{\max}$ and $Q_{i,q,t}\leq Q_{i,q}^{\max}$ for all $i$, $q$ and $t$. Then, all energy demands in the queue $Q_{i,q,t}$ would be served with a maximum queueing delay $R_{i,q}^{\max}$ slots, where
\begin{align} \label{f_22}
R_{i,q}^{\max}\triangleq \lceil(H_{i,q}^{\max}+Q_{i,q}^{\max})/\varepsilon_{i,q}\rceil.
\end{align}
\begin{IEEEproof}
See Appendix A. In addition, in Section \ref{s5}, it can be proved that the constants $H_{i,q}^{\max}$ and $Q_{i,q}^{\max}$ indeed exist.
\end{IEEEproof}

According to the framework of LOT, solving \textbf{P2} is equivalent to solving \textbf{P2'} as follows,
\begin{subequations}\label{f_23}
\begin{align}
(\textbf{P2'})~~&\min~\mathop {\lim\sup}\limits_{T \to \infty}\frac{1}{T}\sum\limits_{t=0}^{T-1} \mathbb{E}\{\Gamma_t\},  \\
s.t.&~(1),(2),(4)-(6),(9),(10),(12)-(14),(17),(18),\nonumber \\
&\text{Queues}~Q_{i,q,t},~H_{i,q,t},~\text{and}~Z_{i,q,t}~\text{are mean rate stable}.\nonumber
\end{align}
\end{subequations}

\subsection{The Proposed Realtime Algorithm}

Define $\boldsymbol{\Theta_t}\triangleq (\boldsymbol{Q_t},\boldsymbol{H_t},\boldsymbol{Z_t})$ as the concatenated vector of the real workload queue, virtual workload queue and virtual energy queue, where
\begin{align}
&\boldsymbol{Q_t}=(Q_{1,1,t},\cdots,Q_{1,M_1,t},\cdots,Q_{N,1,t},\cdots,Q_{N,M_N,t}), \nonumber \\
&\boldsymbol{H_t}=(H_{1,1,t},\cdots,H_{1,M_1,t},\cdots,Z_{N,1,t},\cdots,Z_{N,M_N,t}), \nonumber \\
&\boldsymbol{Z_t}=(Z_{1,t},Z_{2,t},\cdots,Z_{N,t}). \nonumber
\end{align}

To keep the stability of all queues, we first define a weighted quadratic Lyapunov function as follows,
\begin{align}\label{f_25}
    \mathcal{L}_t \buildrel \Delta \over = \frac{1}{2}\sum\limits_{i = 1}^N \Big(\sum\limits_{q=1}^{M_i}w(Q_{i,q,t}^2+H_{i,q,t}^2)+Z_{i,t}^2\Big),
\end{align}
where $w$ is a positive weight for workload queues, which indicates the relative importance of the workload queues with
respect to the energy queues.

Then, a one-slot conditional Lyapunov drift could be obtained below,
\begin{align}\label{f_26}
    \Delta_t = \mathbb{E}\{ \mathcal{L}_{t+1} - \mathcal{L}_t|\boldsymbol{\Theta_t}\},
\end{align}
where the expectation is taken with respect to the randomness of workloads, renewable generation outputs, electricity prices, and the randomness in control policies.

Next, by adding a function of the expected operational cost in a slot to \eqref{f_26}, we can obtain a \emph{drift-plus-penalty} term as follows,
\begin{align}\label{f_27}
\Delta V_t =\Delta_t + V\mathbb{E}\{\Gamma_t|\boldsymbol{\Theta_t}\}.
\end{align}

\textbf{\emph{Lemma 2~(Drift Bound)}}
The \emph{drift-plus-penalty} term satisfies the following inequality for all slots,
\begin{align} \label{f_28}
\Delta V_t\leq &\Omega_0+V\mathbb{E}\{\Gamma_{t}|\boldsymbol{\Theta_t}\} \nonumber \\
&+\mathbb{E}\{\sum\limits_{i=1}^{N}\sum\limits_{q=1}^{M_i}wQ_{i,q,t}(\pi_{i,q,t}-x_{i,q,t})|\boldsymbol{\Theta_t}\} \nonumber \\
&+\mathbb{E}\{\sum\limits_{i=1}^{N}\sum\limits_{q=1}^{M_i}wH_{i,q,t}(\varepsilon_{i,q}-x_{i,q,t})|\boldsymbol{\Theta_t}\}\nonumber \\
&+\mathbb{E}\{\sum\limits_{i=1}^{N}Z_{i,t}(\eta_{c,i} u_{c,i,t}-\frac{1}{\eta_{d,i}}u_{d,i,t})|\boldsymbol{\Theta_t}\}.
\end{align}
where $\Omega_0$ is given by
\begin{align}
\Omega_0=&\sum\limits_{i=1}^{N}\sum\limits_{q=1}^{M_i}\Bigg(w\frac{(\pi_{i,q}^{\max})^2+(x_{i,q}^{\max})^2+\max\{\varepsilon_{i,q}^2,(x_{i,q}^{\max})^2\}}{2}\Bigg)\nonumber\\
&+\sum\limits_{i=1}^{N}\frac{\max\{(\eta_{c,i} u_{i,\text{cmax}})^2,(\frac{1}{\eta_{d,i}}u_{i,\text{dmax}})^2\}}{2}.
\end{align}

\begin{IEEEproof}
See Appendix B.
\end{IEEEproof}

Minimizing the R.H.S. of the upper bound of drift-plus-penalty term in each slot $t$, we have the following optimization problem \textbf{P3} as follows,
\begin{subequations}\label{f_29}
\begin{align}
(\textbf{P3})~&\min~V\Gamma_{t}-\sum\limits_{i=1}^{N}\sum\limits_{q=1}^{M_i}w\big(Q_{i,q,t}+H_{i,q,t}\big)x_{i,q,t}+\nonumber  \\
&~~~~~~~~\sum\limits_{i=1}^{N}Z_{i,t}(\eta_{c,i} u_{c,i,t}-\frac{1}{\eta_{d,i}}u_{d,i,t}) \\
s.t.&~(1),(2),(4)-(6),(9),(10),(12)-(14),(17),(18). \nonumber
\end{align}
\end{subequations}

Since \textbf{P3} neglects the constraint \eqref{f_9}, we can obtain \textbf{P4} by adding \eqref{f_9} into the constraints of \textbf{P3}, i.e.,
\begin{subequations}\label{f_30}
\begin{align}
(\textbf{P4})~&\min~V\Gamma_{t}-\sum\limits_{i=1}^{N}\sum\limits_{q=1}^{M_i}w\big(Q_{i,q,t}+H_{i,q,t}\big)x_{i,q,t}+\nonumber  \\
&~~~~~~~~\sum\limits_{i=1}^{N}Z_{i,t}(\eta_{c,i} u_{c,i,t}-\frac{1}{\eta_{d,i}}u_{d,i,t}) \\
s.t.&~(1),(2),(4)-(6),(9)-(14),(17)-(18). \nonumber
\end{align}
\end{subequations}

Since the constraint \eqref{f_11} is nonlinear, \textbf{P4} is a nonlinear programming problem. To simplify the computation, we can first ignore the nonlinear constraint \eqref{f_11}, and then adjust the obtained solution to satisfy \eqref{f_11}. Based on the above description, an algorithm for \textbf{P1} could be described by Algorithm 1, where \textbf{P5} is defined as follows,

\begin{subequations}\label{f_31}
\begin{align}
(\textbf{P5})~&\min~V\Gamma_{t}-\sum\limits_{i=1}^{N}\sum\limits_{q=1}^{M_i}w\big(Q_{i,q,t}+H_{i,q,t}\big)x_{i,q,t}+\nonumber  \\
&~~~~~~~~\sum\limits_{i=1}^{N}Z_{i,t}(\eta_{c,i} u_{c,i,t}-\frac{1}{\eta_{d,i}}u_{d,i,t}) \\
s.t.&~(1),(2),(4)-(6),(9)-(13),(17)-(18).
\end{align}
\end{subequations}

\begin{algorithm}[h]
\caption{: Realtime Algorithm for Operational Cost Minimization Problem}
\begin{algorithmic}[1]
\STATE \textbf{For} each slot $t$ \textbf{do}\\
\STATE Observing system states at the starting point of time slot $t$: $Q_{i,q,t}$, $H_{i,q,t}$, $Z_{i,t}$, $\lambda_{f,t}$, $\pi_{i,q,t}$, $r_{i,t}$, $X_{i,t}$ and $W_{i,t}$; \\
\STATE Choose control decisions $d_{f,i,t}$, $x_{i,q,t}$, $c_{i,t}$, $u_{c,i,t}$, $u_{d,i,t}$, $e_{i,q,t}$, $g_{i,t}$, as the solution to $\textbf{P5}$; \\
\STATE Generate a new solution based on the following equations so that the constraint \eqref{f_11} could be satisfied: $\hat{u}_{c,i,t}=\max\{u_{c,i,t}-\frac{1}{\eta_{c,i}\eta_{d,i}}u_{d,i,t},0\}$, $\hat{u}_{d,i,t}=\max\{u_{d,i,t}-\eta_{c,i}\eta_{d,i}u_{c,i,t},0\}$, $\hat{d}_{f,i,t}=d_{f,i,t}$, $\hat{g}_{i,t}=g_{i,t}$, $\hat{e}_{i,q,t}=e_{i,q,t}$, $\hat{x}_{i,q,t}$=$x_{i,q,t}$, $\hat{c}_{i,t}=c_{i,t}+(\hat{u}_{c,i,t}-u_{c,i,t})+(u_{d,i,t}-\hat{u}_{d,i,t})$.\\
\STATE Updating $Q_{i,q,t}$, $H_{i,q,t}$, $Z_{i,t}$ with the new solution according to (3),~(25),~and~(24).
\STATE \textbf{End}
\end{algorithmic}
\end{algorithm}

\textbf{Remarks}: Note that the constraints \eqref{f_5},~\eqref{f_6}, and \eqref{f_12} in \textbf{P1} are not considered in Algorithm 1, the solution generated by Algorithm 1 may be infeasible to \textbf{P1}. In Section~\ref{s5}, we will show that Algorithm 1 can guarantee the feasibilities of \eqref{f_5},~\eqref{f_6}, and \eqref{f_12}.

\subsection{Distributed Implementation}
To solve \textbf{P5} efficiently, we propose a distributed implementation for the proposed realtime algorithm. A possible way of obtaining a distributed algorithm for \textbf{P5} is based on dual decomposition, which decomposes the Lagrangian dual problem of \textbf{P5} into independent subproblems that could be solved in parallel. Unfortunately, the objective function in \textbf{P5} is not strictly convex since $\Gamma_{2,t}$ and $\Gamma_{3,t}$ are linear functions. As a result, dual decomposition cannot be applied, for otherwise the Lagrangian is unbounded below\cite{Boyd2011}. Since ADMM could be used to solve a large-scale convex optimization problem without assuming strict convexity of the separable objective function, we are thus motivated to design a ADMM-based distributed algorithm.

In order to utilize the ADMM framework, \textbf{P5} is transformed into the following problem equivalently.
\begin{subequations}\label{f_37}
\begin{align}
(\textbf{P6})~&\min~V\Gamma_{t}-\sum\limits_{i=1}^{N}\sum\limits_{q=1}^{M_i}w\big(Q_{i,q,t}+H_{i,q,t}\big)b_{i,q,t}+\nonumber  \\
&~~~~~~~~\sum\limits_{i=1}^{N}Z_{i,t}(\eta_{c,i} u_{c,i,t}-\frac{1}{\eta_{d,i}}u_{d,i,t}) \\
s.t.&~(1),(2),(4),(10)-(13),(18),\\
&\sum\limits_{f=1}^F a_{f,i,t}+\sum\limits_{q=1}^{M_i}(b_{i,q,t}-e_{i,q,t})+h_i = C_i,\\
&g_{i,t}+c_{i,t}+u_{d,i,t}-u_{c,i,t}+\beta_i h_i=m_i,\\
&e_{i,q,t}+z_{i,q}=b_{i,q,t}, \\
&d_{f,i,t}=a_{f,i,t}, \\
&x_{i,q,t}=b_{i,q,t},
\end{align}
\end{subequations}
where $h_i$ and $z_{i,q}$ are a set of nonnegative slack variables; $a_{f,i,t}$ and $b_{i,q,t}$ are nonnegative auxiliary variables; the constant $m_i=\alpha_i+\beta_iC_i-r_{i,t}$; the decision variables are $d_{f,i,t},a_{i,q,t},x_{i,q,t},b_{i,q,t},e_{i,q,t},c_{i,t},g_{i,t}, u_{c,i,t},u_{d,i,t},h_i,z_{i,q}$.

If ADMM framework applies to \textbf{P6} directly, eleven blocks would be generated since there are eleven kinds of variables. For ADMM with more than two blocks, the convergence is still an open question. In this paper, we adopt the algorithm in \cite{He2012} to solve \textbf{P6}, which is called as ADM-G (ADM with Gaussian back substitution). The global convergence of ADM-G is provable under mild assumptions. Following the method in our previous work\cite{IoTLiangYu2016}\cite{Liang2016TSG}, it is easy to check that ADM-G framework could result in an optimal solution of \textbf{P6} if the optimal solution is non-empty. Due to the space limit, we omit the proof for simplicity. Following the framework of ADM-G, we can obtain a distributed implementation of the proposed realtime algorithm in Appendix C.

\section{Algorithmic Performance Analysis}\label{s4}
In this section, we provide the performance analysis of the designed distributed realtime algorithm. Specifically, we first present a Lemma, which offers a sufficient condition for the charging and discharging of the ESS in SMG $i$ at slot $t$ under the proposed algorithm. Then, based on the Lemma, a Theorem is proposed to show the feasibility of the Algorithm 1 for \textbf{P1}.

\textbf{\emph{Lemma 3.}}
Define $\gamma_{i,\min}$=$\min\{X_{i,\min},W_{i,\min},A_{i,\min}'\}$. Then,
\begin{enumerate}
  \item If $Z_{i,t}<-V\eta_{d,i}\gamma_{i,\max}$, the optimal discharging decision is $u_{d,i,t}^{*}=0$,
  \item If $Z_{i,t}>-\frac{V}{\eta_{c,i}}\gamma_{i,\min}$, the optimal charging decision is $u_{c,i,t}^{*}=0$.
\end{enumerate}

\begin{IEEEproof}
See Appendix D.
\end{IEEEproof}

With the above lemma, a theorem is provided to show the performance of the designed algorithm.

\textbf{\emph{Theorem 1}}
Suppose $x_{i,q}^{\max}\geq \max[\pi_{i,q}^{\max},~\varepsilon_{i,q}]$. If $Q_{i,q,0}=H_{i,q,0}=0$, the proposed algorithm can provide the following guarantees:
\begin{enumerate}
  \item The queues $Q_{i,q,t}$ and $H_{i,q,t}$ are bounded by $Q_{i,q}^{\max}$ and $H_{i,q}^{\max}$, respectively. In particular, $Q_{i,q}^{\max}=V\beta_iX_{i}^{\max}/w+\pi_{i,q}^{\max}, H_{i,q}^{\max}=V\beta_iX_{i}^{\max}/w+\varepsilon_{i,q}$.
  \item The maximum queueing delay $R_{i,q}^{\max}=\left\lceil\frac{2V\beta_iX_{i}^{\max}/w+\pi_{i,q}^{\max}+\varepsilon_{i,q}}{\varepsilon_{i,q}}\right\rceil.$
\item The energy queue $D_{i,t}$ satisfies the following for all time slot $t$: $D_{i,\min}\leq D_{i,t}\leq D_{i,\max}$.
  \item The solution of the proposed algorithm is feasible to the original problem \textbf{P1}.
  \item Compared with the optimal solution of \textbf{P3}, the maximum optimality loss due to the incorporation of ramping constraints in \textbf{P4} is $\Omega_1=\sum\nolimits_{i=1}^N V(1-\epsilon_i)c_{i,\max}\gamma_{i,\max}$.
  \item Compared with the optimal solution of \textbf{P5}, the maximum optimality loss in the aspect of $\Gamma_t$ caused by the online solution adjustment is $\Omega_2=\sum\nolimits_{i=1}^N\big(\sigma_i(u_{i,\text{cmax}}^2+u_{i,\text{dmax}}^2)+\delta_{1,i}c_{i,\max}^2+\delta_{2,i}c_{i,\max}\big)$.
  \item If $\varepsilon_{i,q}\leq \mathbb{E}\{\pi_{i,q,t}\}$ and the uncertain parameters $\lambda_{f,t}$, $\pi_{i,q,t}$, $r_{i,t}$, $X_{i,t}$ and $W_{i,t}$ are i.i.d. over slots, the proposed algorithm offers the following performance guarantee, i.e.,
      $\mathop {\lim\sup}\limits_{T \to \infty}\frac{1}{T}\sum\nolimits_{t=0}^{T-1} \mathbb{E}\{\Gamma_{t}\}\leq y_1+\Omega_2+\frac{\Omega_0+\Omega_1}{V}$, where $y_1$ is the optimal objective value of \textbf{P1}.
\end{enumerate}

\begin{IEEEproof}
See Appendix E.
\end{IEEEproof}

\section{Performance Evaluation} \label{s5}
\subsection{Simulation Setup}\label{s71}
We intend to evaluate the performance of the proposed algorithm in six months with 4320 1-hour slots. To model the generation cost of conventional generator $i$, a quadratic polynomial is adopted as in \cite{ZhangRui2015}, i.e., $A_i(c_{i,t})=\delta_{1,i} c_{i,t}^2+\delta_{2,i}c_{i,t}+\delta_{3,i}$. For simplicity, we set $\delta_{1,i}=\delta_{3,i}=0$, $\delta_{2,i}=273\$/MW$\cite{Gen2014}. To model the battery depreciation cost, a function is considered as in \cite{Rivera2013}, i.e., $B_i(u_{c,i,t},u_{d,i,t})=\sigma_{i}(u_{c,i,t}^2+u_{d,i,t}^2)$. We set $\epsilon_i=1$, $\sigma_{i}=100$, $\eta_{c,i}=\eta_{d,i}=1$. The parameters associated with data centers and front-end servers are given as follows, i.e., $F=1$, $N=3$, $M_1=40000$, $M_2=30000$, $M_3=30000$, $P_{i,\text{peak}}=200$ Watts, $P_{i,\text{idle}}=140$ Watts, $PUE_1=1.1$, $PUE_2=1.2$, $PUE_3=1.3$. $u_{i,\text{cmax}}=u_{i,\text{dmax}}=0.5$ MW\cite{liangyu2015}. $\omega=1\times 10^{-4}$\cite{Xu2013}, $\theta_i$=0.1, $V=V^{\max}$, $\varepsilon_{i,q}=(2V\beta_iX_{i}^{\max}/w+\pi_{i,q}^{\max})/(\mathcal{T}_{i,q}-1)$, $x_{i,q}^{\max}=\pi_{i,q}^{\max}$. $D_{1,\max}=8.8$ MWh, $D_{2,\max}=7.2$ MWh, $D_{3,\max}=7.8$ MWh (i.e., data centers could be supported by these ESSs for one hour). In addition, real-world workload traces\footnote{http://ita.ee.lbl.gov/html/traces.html.} and dynamic electricity price traces\footnote{www.nyiso.com; http://www.ercot.com; http://www.pjm.com;} are adopted in simulations. $W_{i,t}=0.9X_{i,t}$\cite{Zhangyu2013}. Suppose that there are two types of batch workloads, i.e., $M_i=2$. To evaluate the impacts of tolerant service delays on the cost reduction under the proposed algorithm, two cases are considered, i.e., case1: $\mathcal{T}_{i,q}\in \{4,8\}$; case2: $\mathcal{T}_{i,q}\in \{12,24\}$. To model the batch workload with type $q$ at data center $i$, we assume that it follows a uniform distribution with parameters 0 and $C_i/(5M_i)$.

To show the advantages of the proposed distributed realtime algorithm, three baselines are adopted.
\begin{itemize}
  \item The first baseline (\emph{B1}) intends to minimize the long-term operational cost with the considerations of energy storage and selling electricity, while batch workloads are processed immediately without delays.
  \item The second baseline (\emph{B2}) intends to minimize the current operational cost considering selling electricity. Moreover, batch workloads are processed immediately. In addition, no energy storage is considered in \emph{B2}.
  \item The three baseline (\emph{B3}) intends to minimize the current operational cost without considering energy storage and selling electricity. Moreover, batch workloads are processed immediately.
\end{itemize}

For simplicity, \emph{Proposed-1} and \emph{Proposed-2} are adopted to denote the performance of the proposed algorithm under case1 and case2, respectively.

\subsection{Simulation Results}\label{s72}

\begin{figure*}
\centering
\subfigure[Maximum queue length (MQL)]{
\begin{minipage}[b]{0.315\textwidth}
\includegraphics[width=1\textwidth]{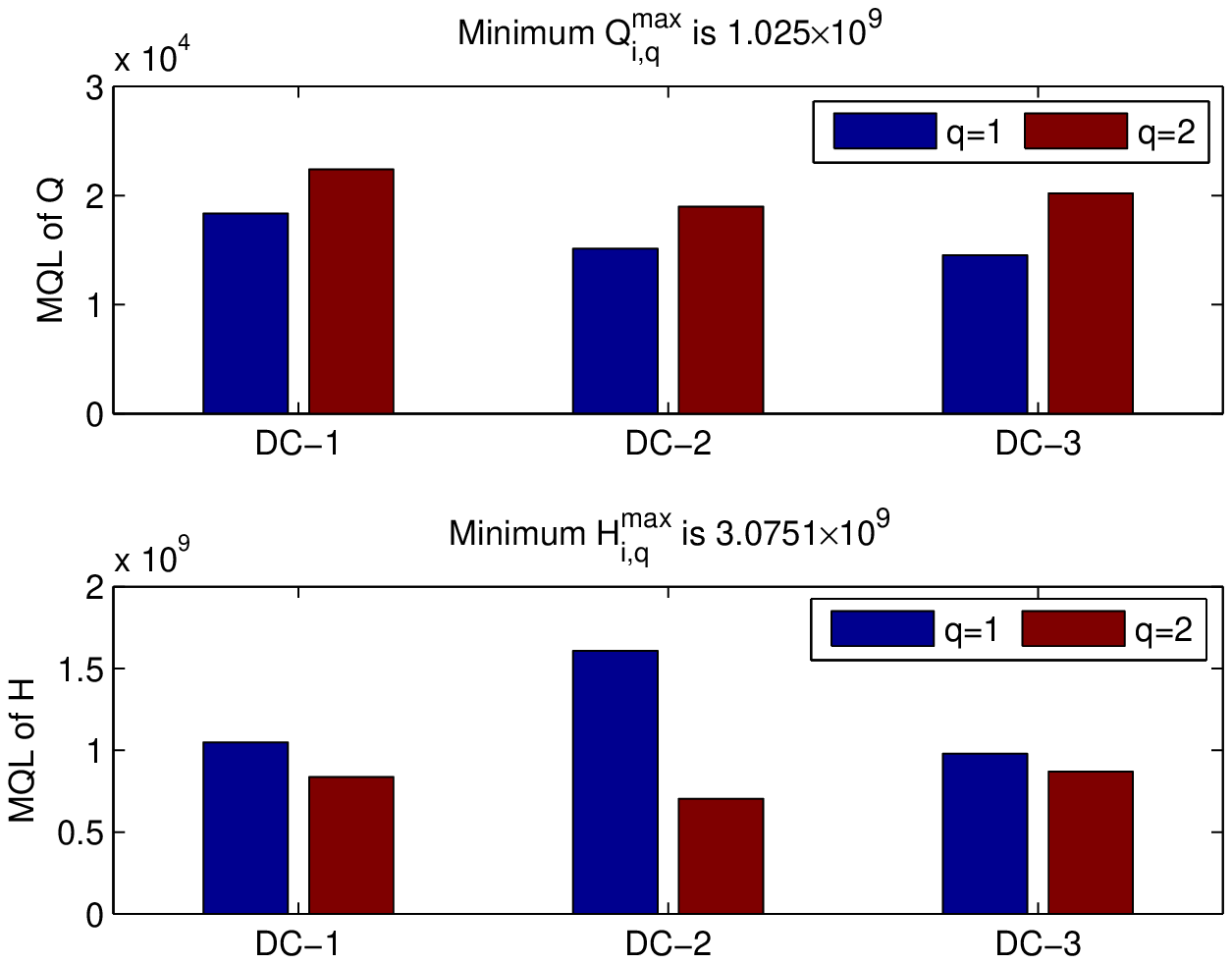}
\end{minipage}
}
\subfigure[Maximum queueing delay (MQD)]{
\begin{minipage}[b]{0.315\textwidth}
\includegraphics[width=1\textwidth]{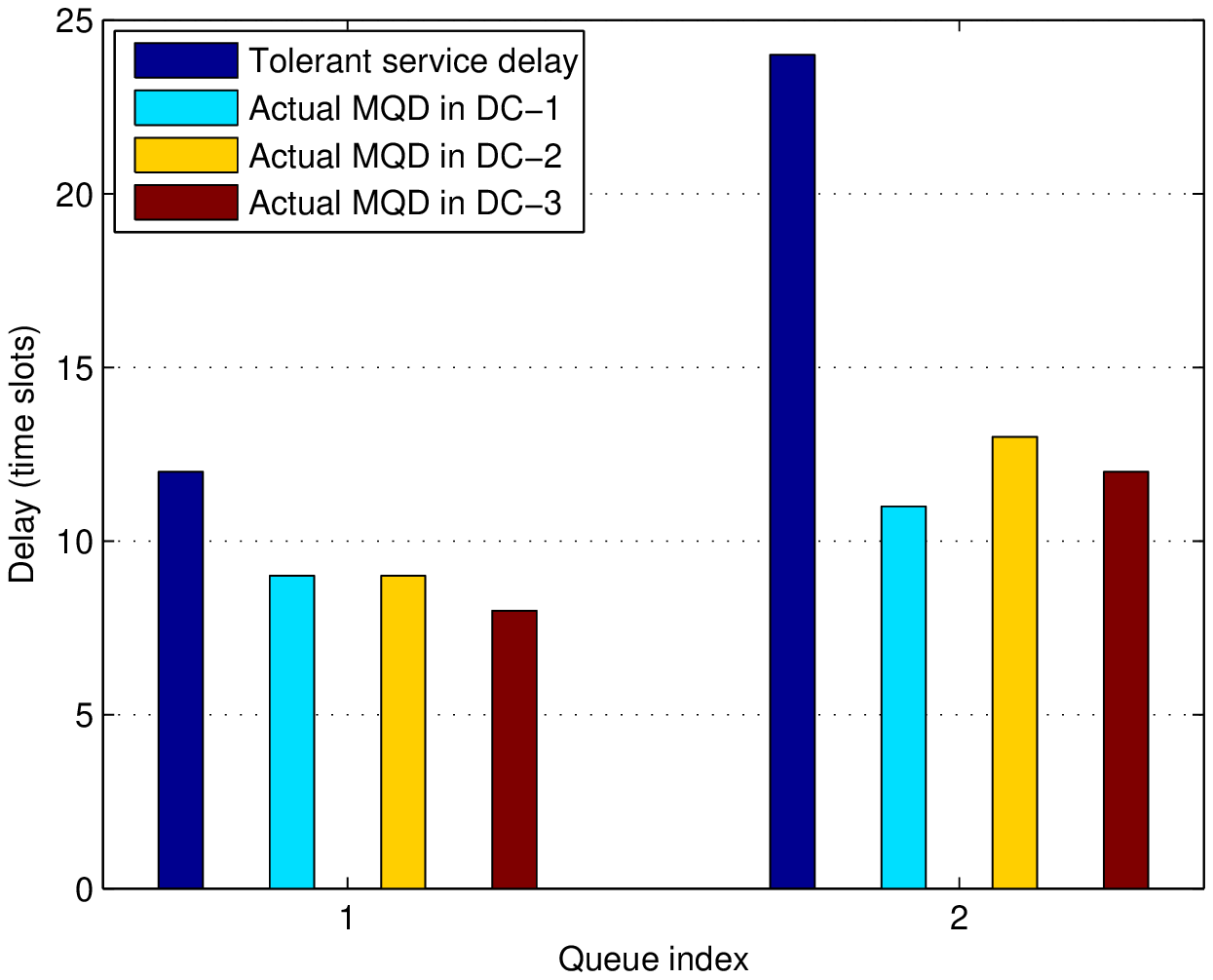}
\end{minipage}
}
\subfigure[Energy level]{
\begin{minipage}[b]{0.315\textwidth}
\includegraphics[width=1\textwidth]{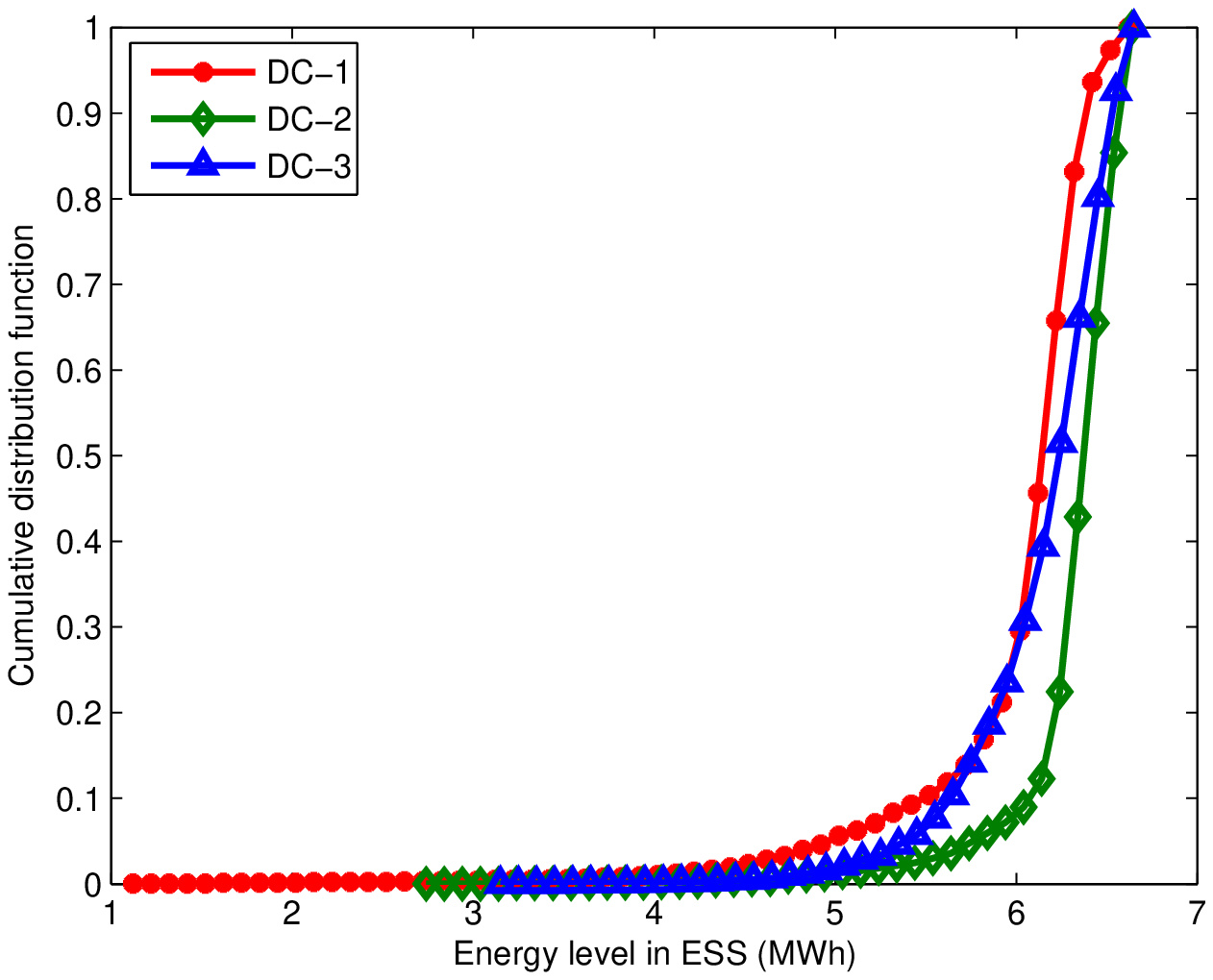}
\end{minipage}
}
\caption{The feasibility of the proposed algorithm} \label{fig_3}
\end{figure*}

\begin{figure*}
\centering
\subfigure[Total cost]{
\begin{minipage}[b]{0.315\textwidth}
\includegraphics[width=1\textwidth]{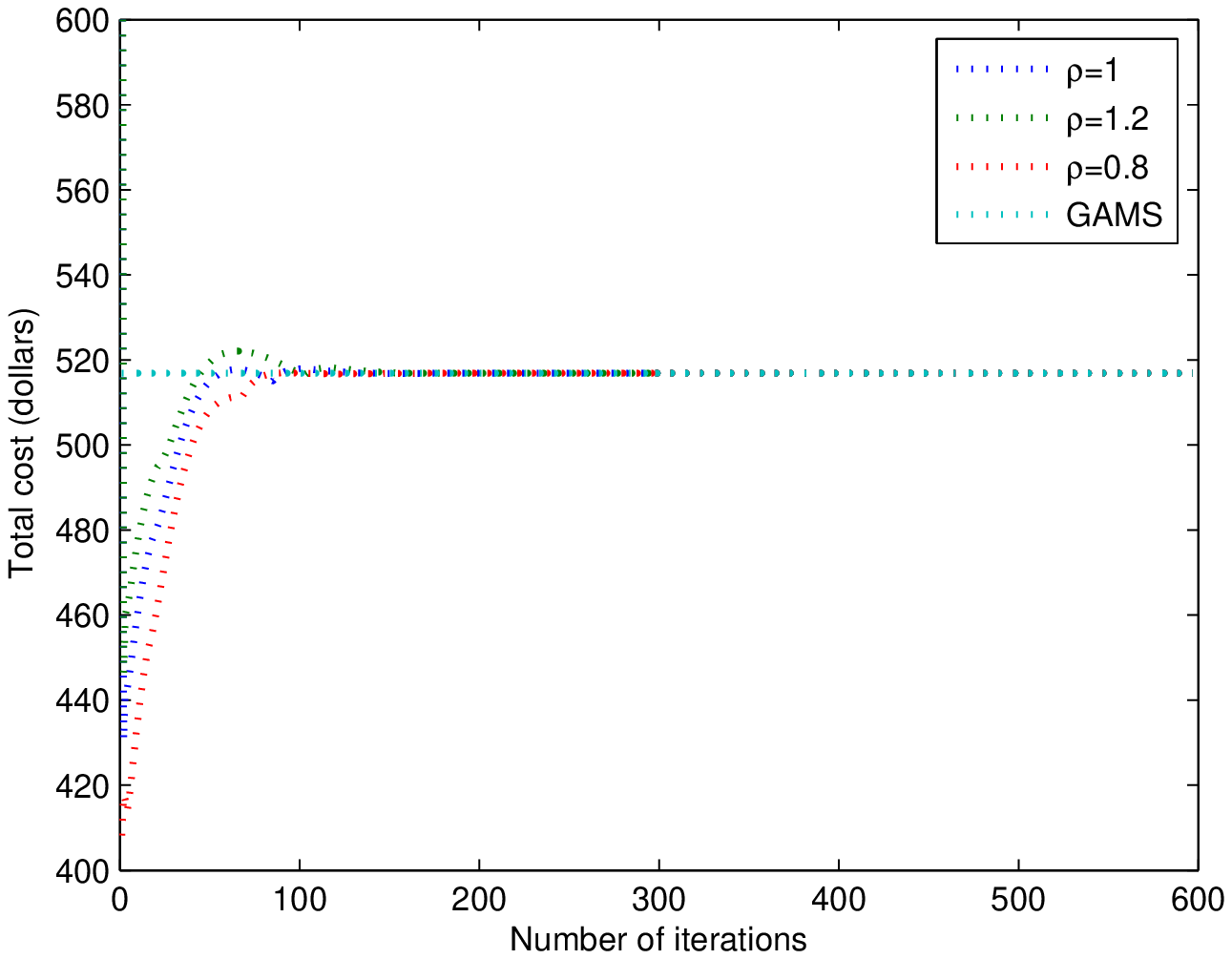}
\end{minipage}
}
\subfigure[Primal residual]{
\begin{minipage}[b]{0.315\textwidth}
\includegraphics[width=1\textwidth]{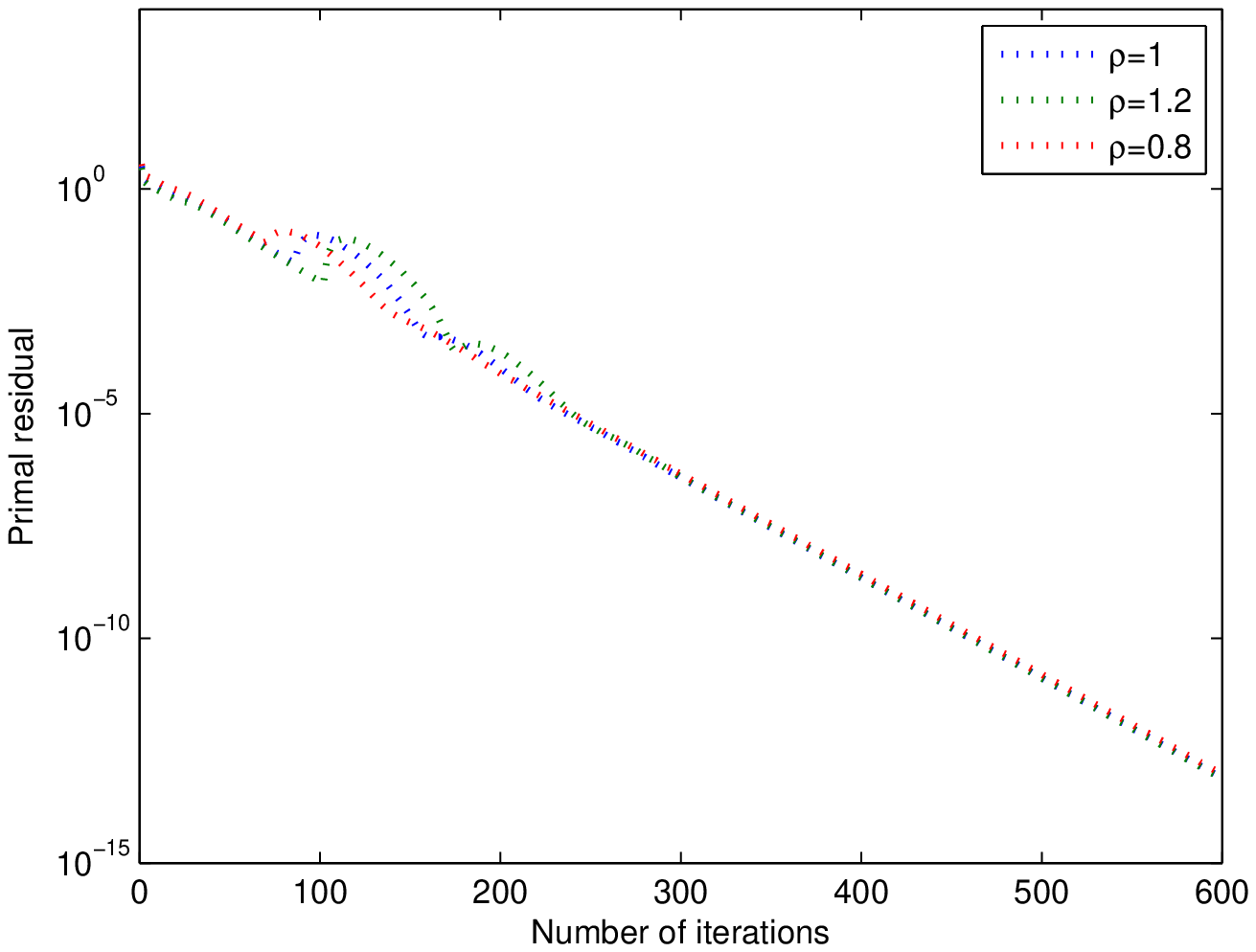}
\end{minipage}
}
\subfigure[Feasibility violation]{
\begin{minipage}[b]{0.315\textwidth}
\includegraphics[width=1\textwidth]{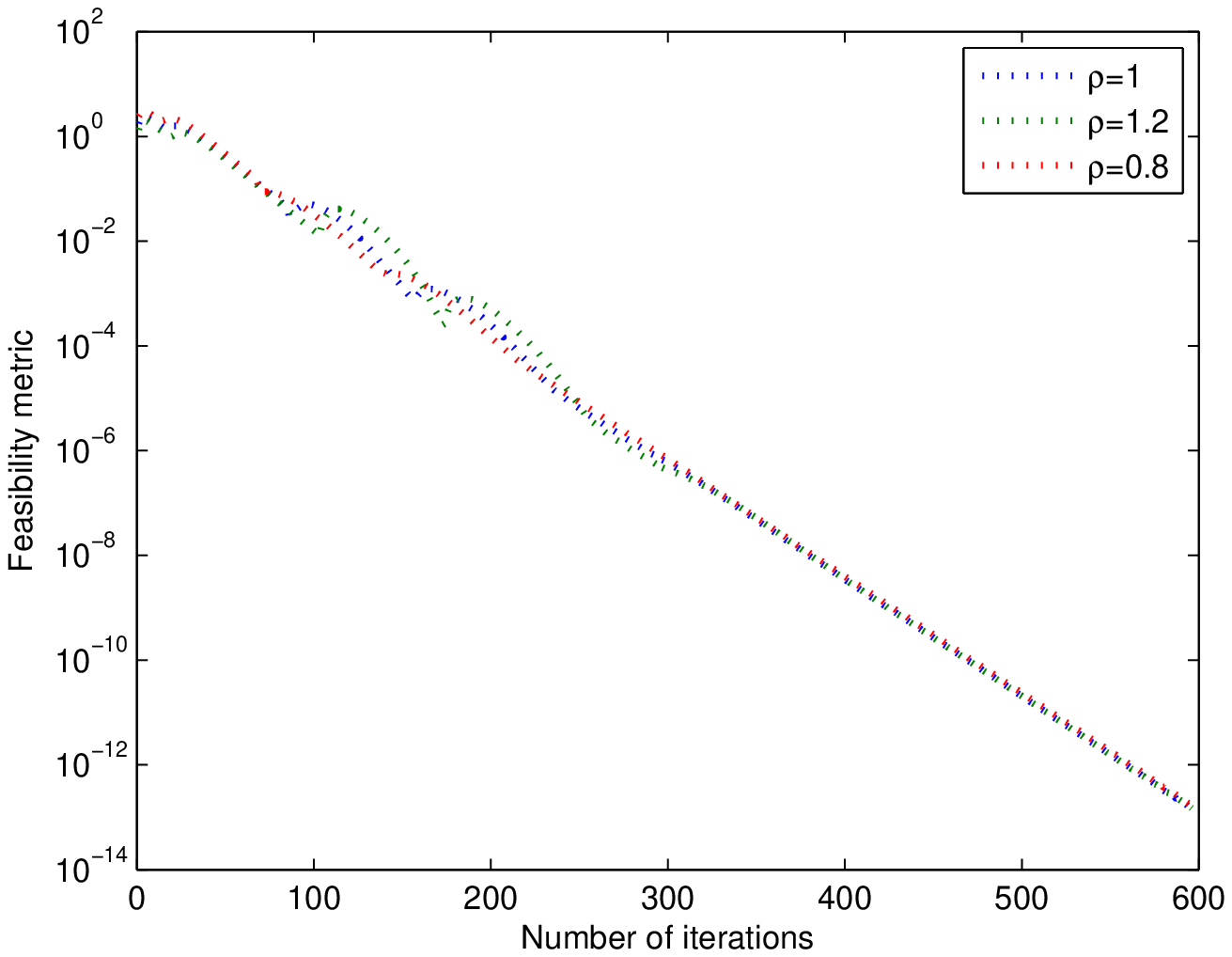}
\end{minipage}
}
\caption{Convergence results of the proposed algorithm} \label{fig_4}
\end{figure*}

\begin{figure*}
\centering
\subfigure[Operational cost]{
\begin{minipage}[b]{0.315\textwidth}
\includegraphics[width=1\textwidth]{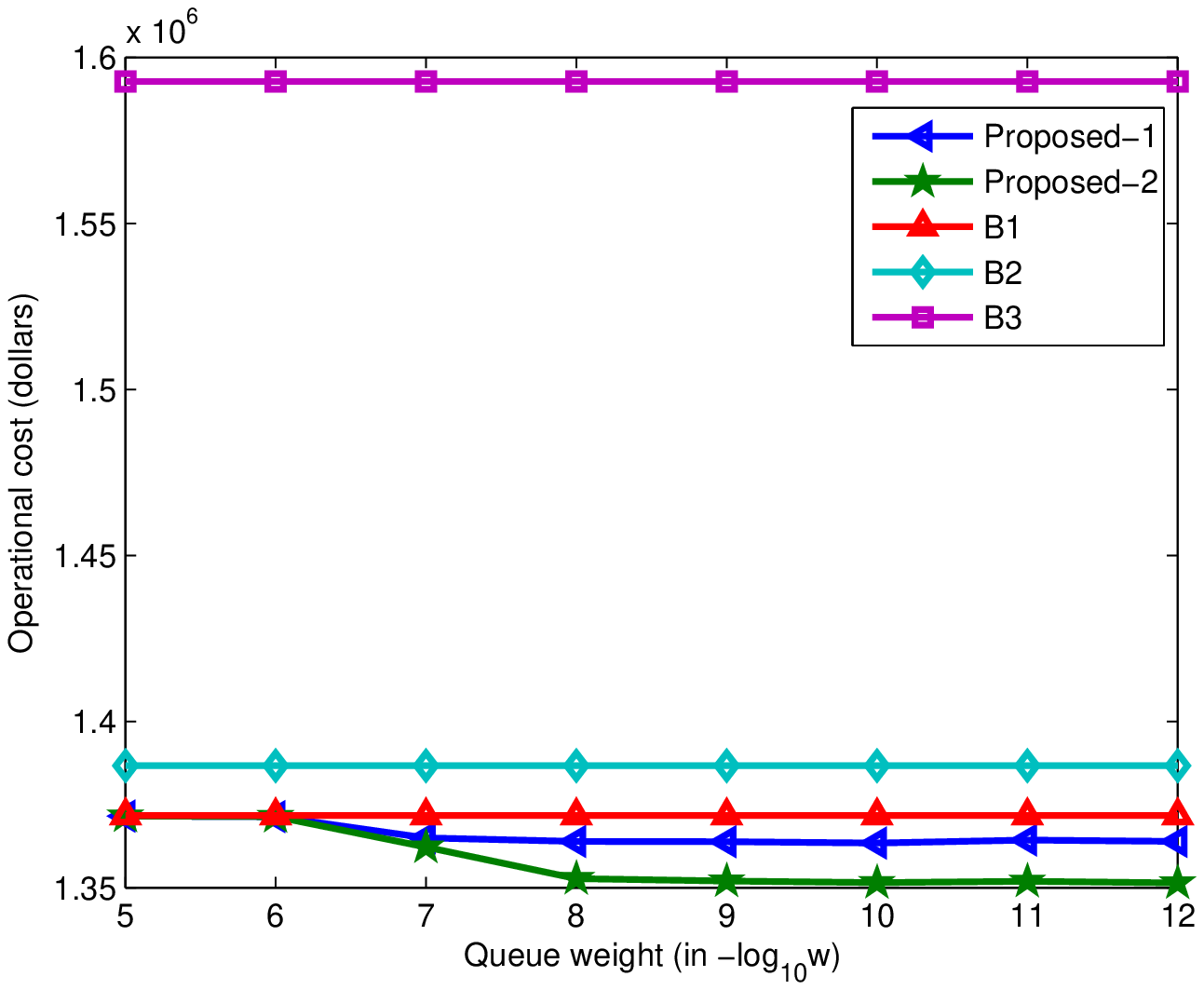}
\end{minipage}
}
\subfigure[Profit of selling electricity]{
\begin{minipage}[b]{0.315\textwidth}
\includegraphics[width=1\textwidth]{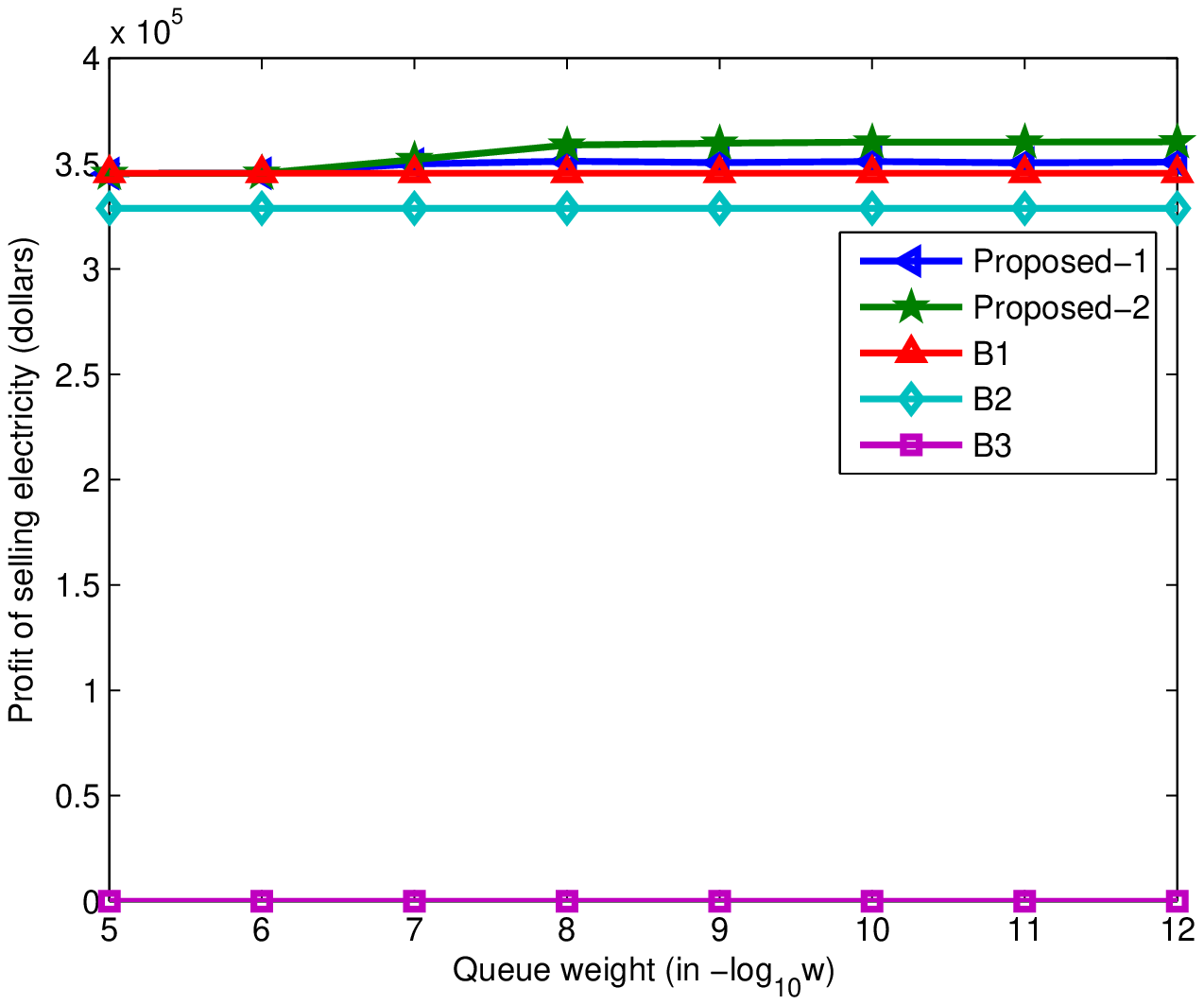}
\end{minipage}
}
\subfigure[AMQD]{
\begin{minipage}[b]{0.315\textwidth}
\includegraphics[width=1\textwidth]{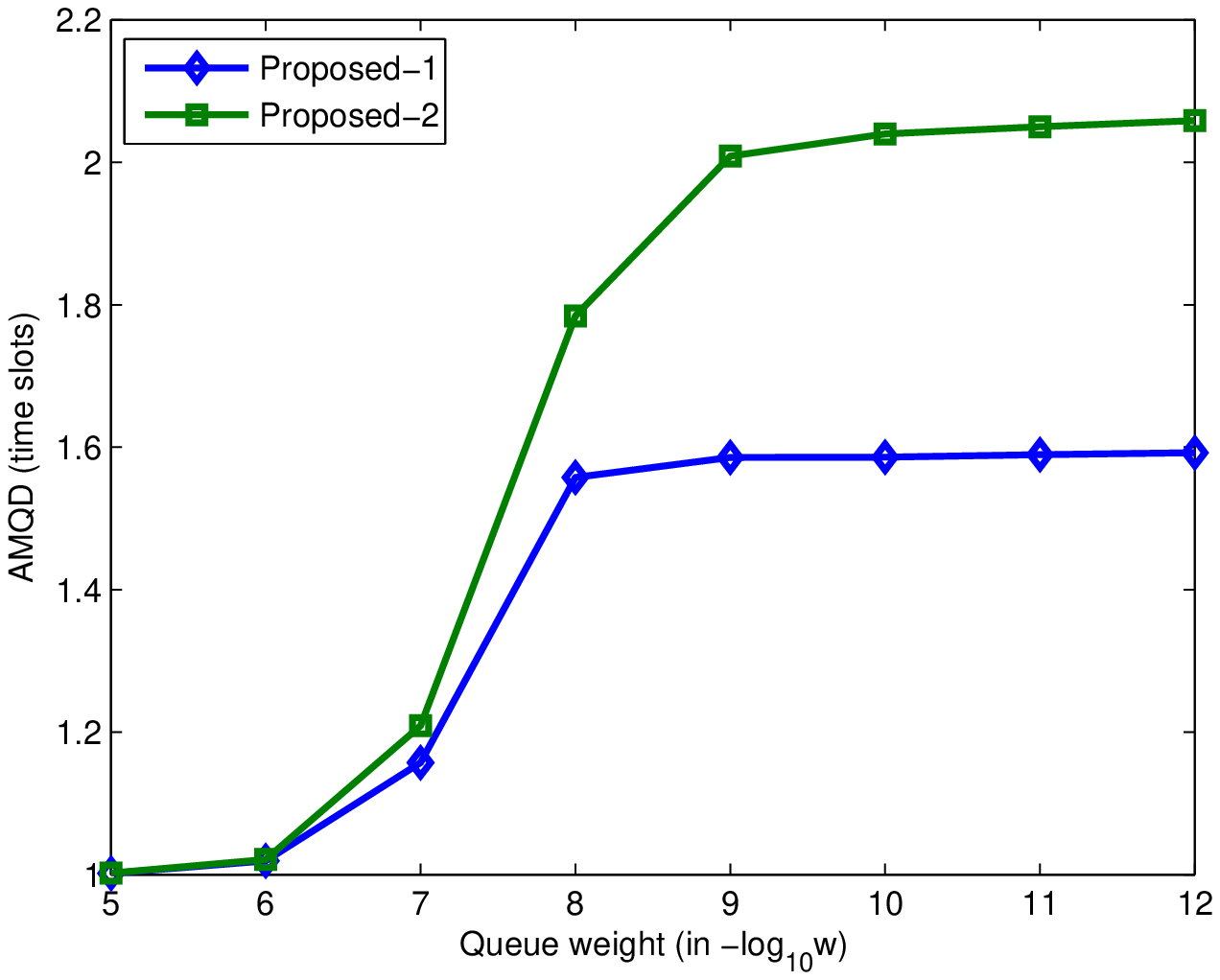}
\end{minipage}
}
\caption{Performances under varying queue weight $w$} \label{fig_5}
\end{figure*}

\begin{figure*}
\centering
\subfigure[Operational cost]{
\begin{minipage}[b]{0.315\textwidth}
\includegraphics[width=1\textwidth]{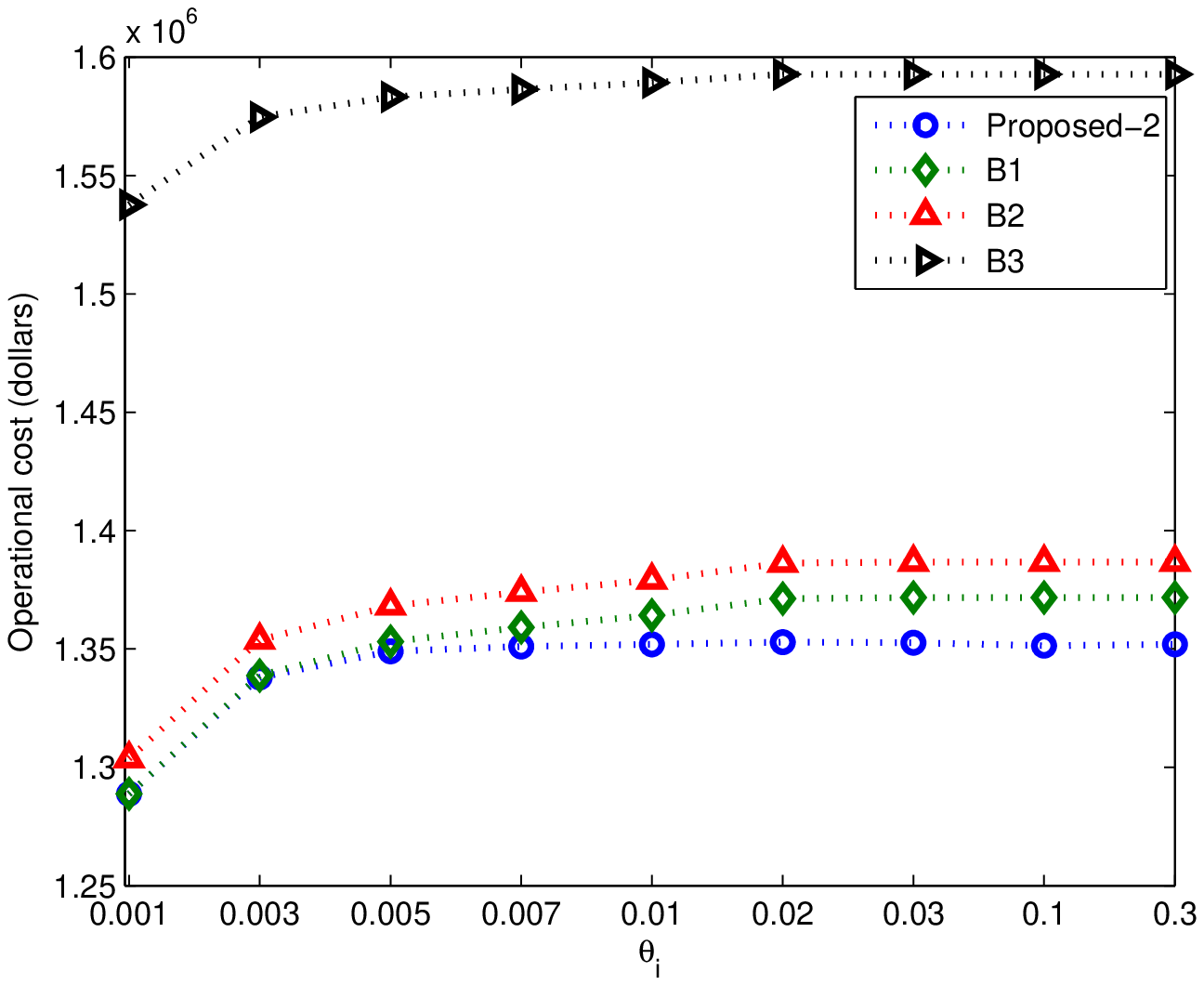}
\end{minipage}
}
\subfigure[Relative cost reduction]{
\begin{minipage}[b]{0.315\textwidth}
\includegraphics[width=1\textwidth]{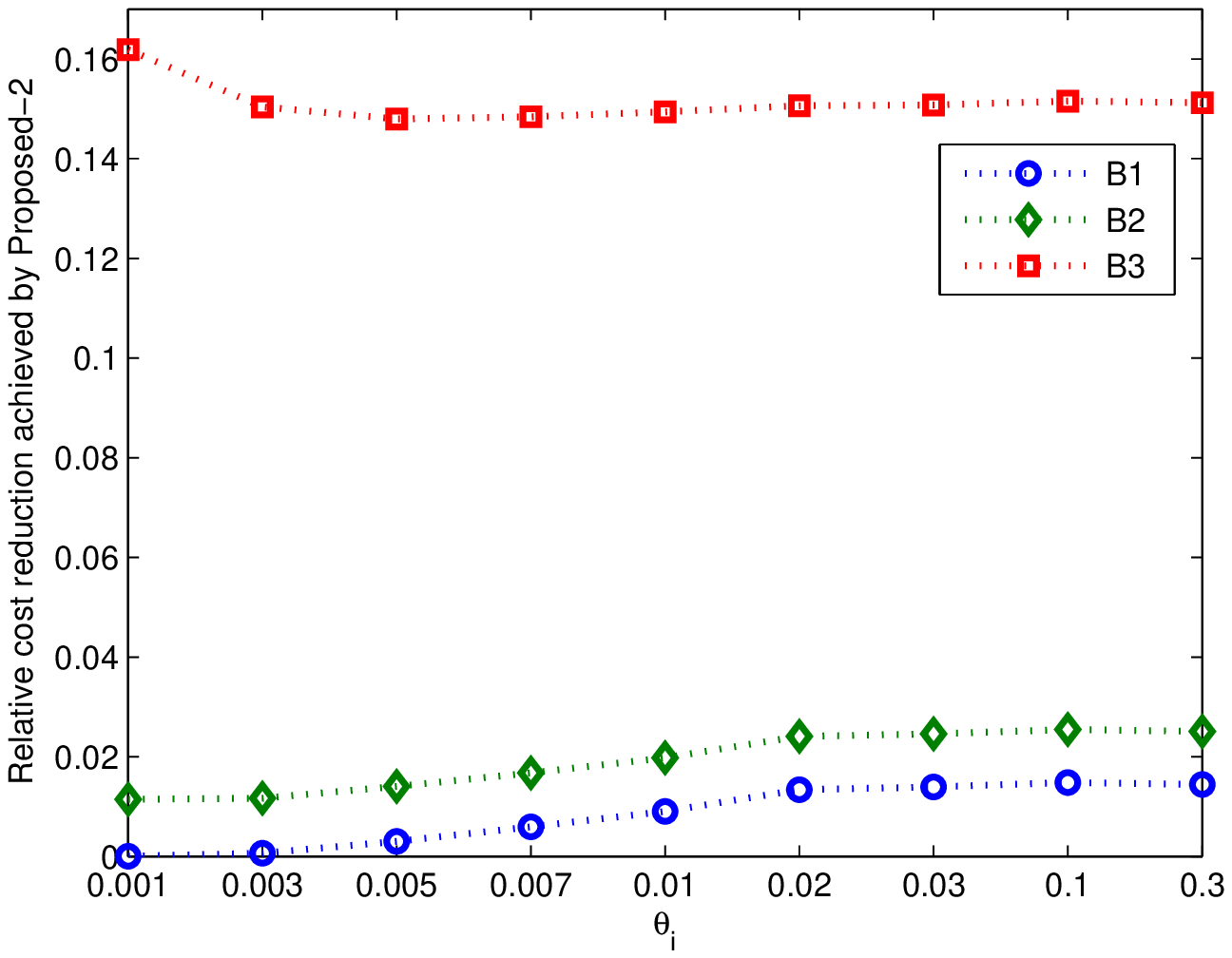}
\end{minipage}
}
\subfigure[Dropping ratio]{
\begin{minipage}[b]{0.315\textwidth}
\includegraphics[width=1\textwidth]{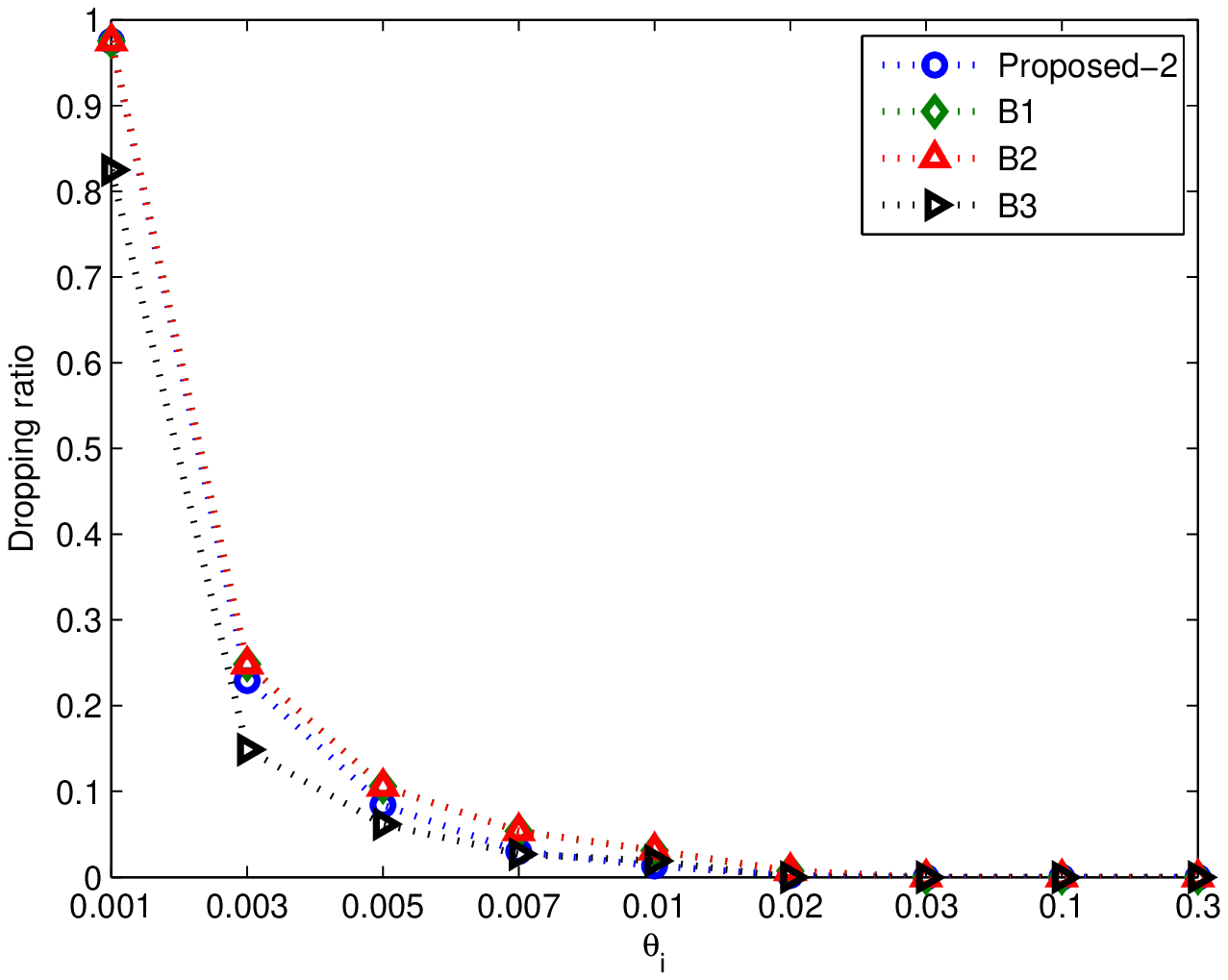}
\end{minipage}
}
\caption{Performances under varying penalty factor $\theta_i$} \label{fig_6}
\end{figure*}

\begin{figure*}
\centering
\subfigure[Operational cost]{
\begin{minipage}[b]{0.315\textwidth}
\includegraphics[width=1\textwidth]{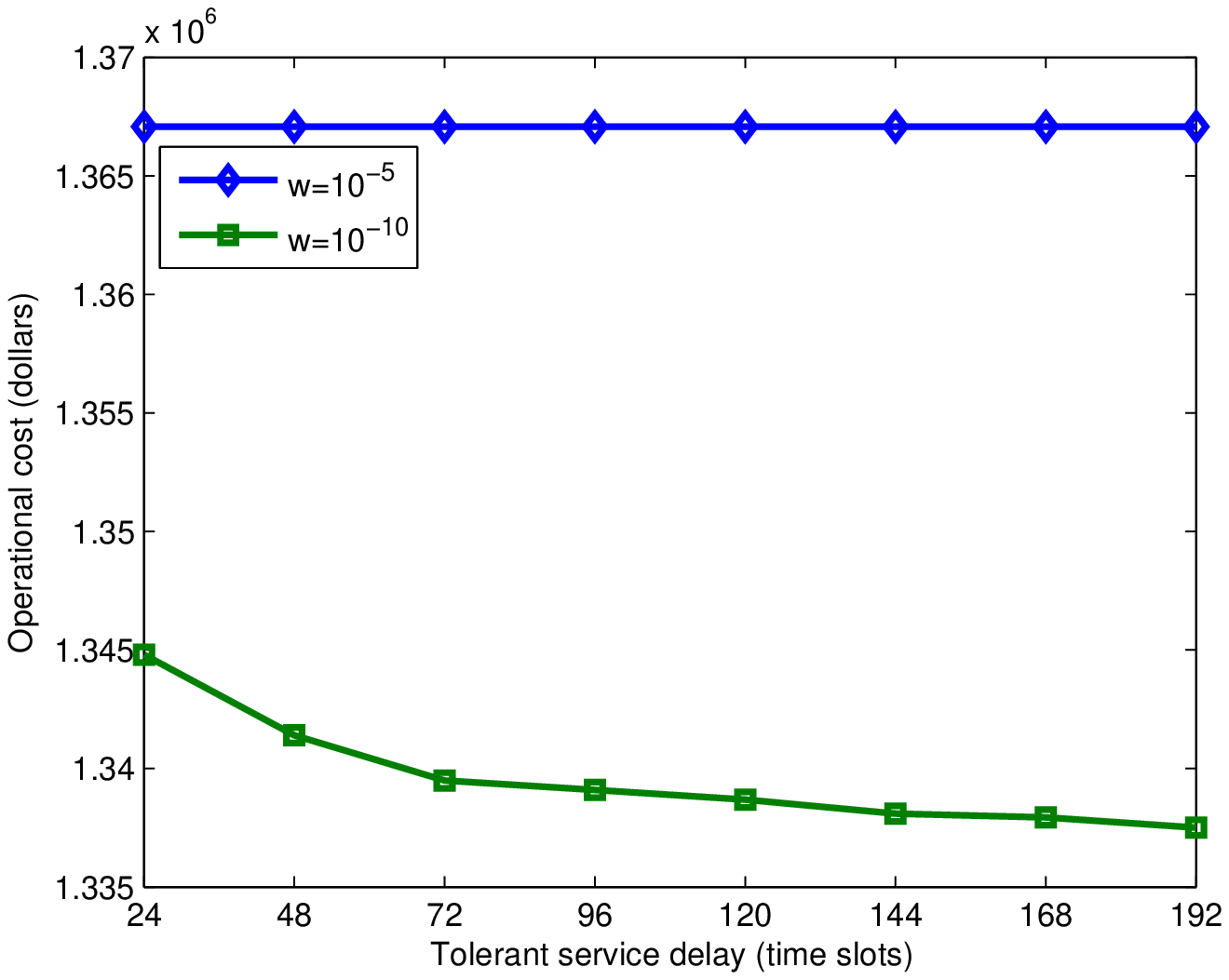}
\end{minipage}
}
\subfigure[Profit of selling electricity]{
\begin{minipage}[b]{0.315\textwidth}
\includegraphics[width=1\textwidth]{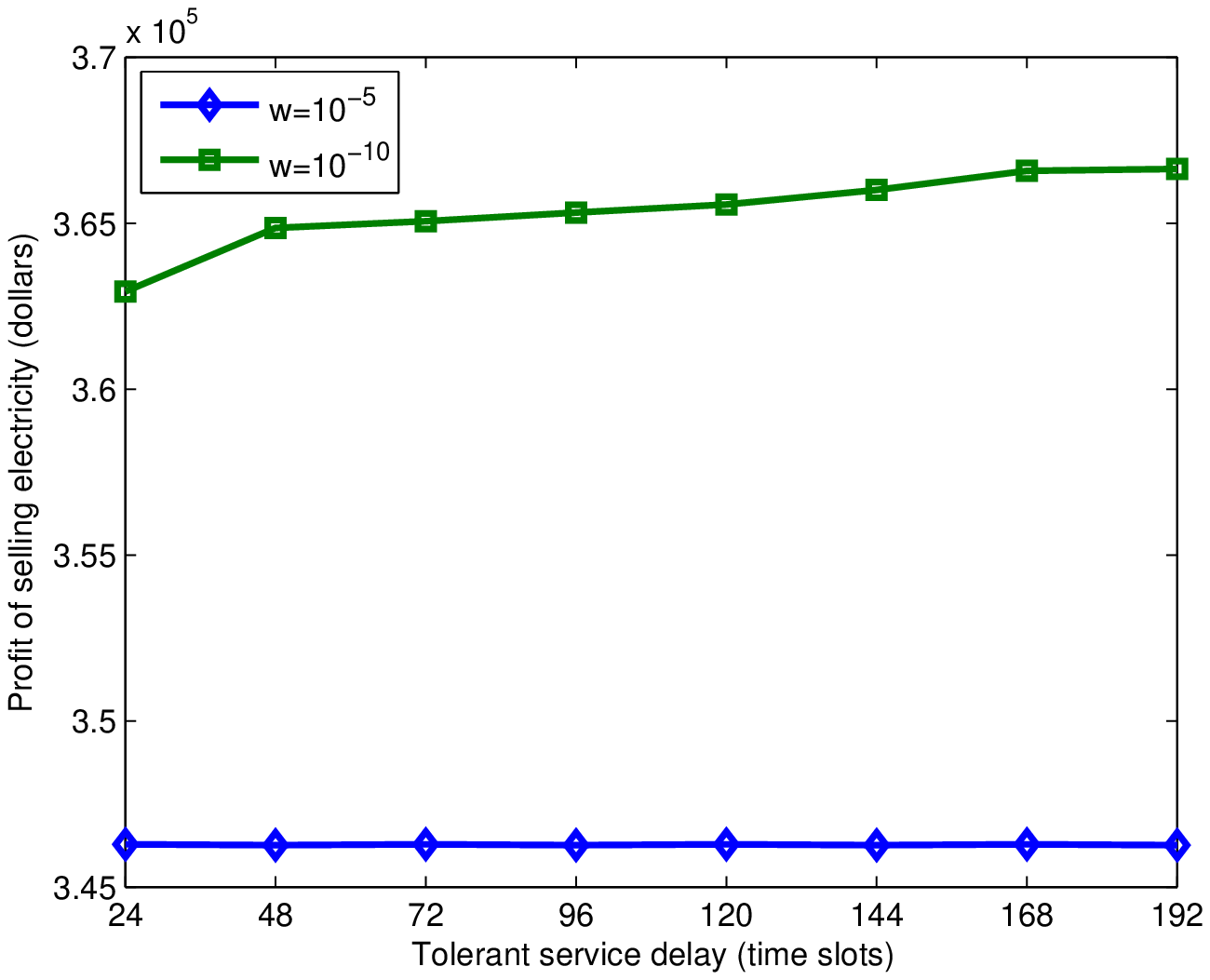}
\end{minipage}
}
\subfigure[AMQD]{
\begin{minipage}[b]{0.315\textwidth}
\includegraphics[width=1\textwidth]{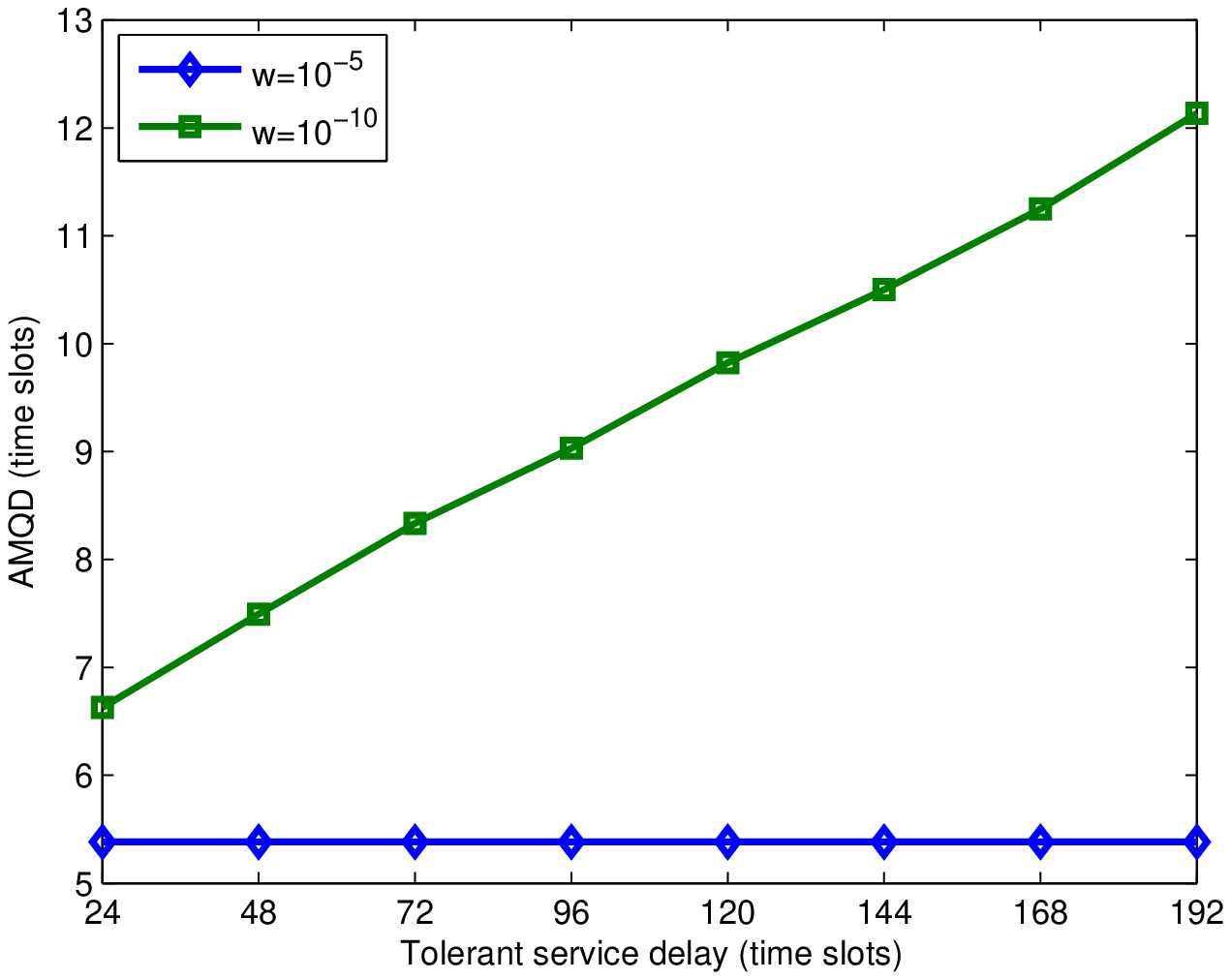}
\end{minipage}
}
\caption{Performances under varying tolerant service delay $\mathcal{T}_{i,q}$} \label{fig_7}
\end{figure*}

\subsubsection{Algorithmic feasibility}
In this subsection, we show the feasibility of the proposed algorithm. Specifically, we need to show that the constraints (7),~(8),~(15) could be satisfied under the proposed algorithm. As indicated in Fig.~\ref{fig_3}~(a), the maximum queue lengths of $Q_{i,q,t}$ and $H_{i,q,t}$ are always smaller than their respective upper bounds (i.e., the constraint (7) holds in all time slots). Moreover, in Fig.~\ref{fig_3}~(b), maximum queueing delays are smaller than the corresponding tolerant service delays, which means that the proposed algorithm could provide the heterogeneous service delay guarantees for all batch workloads, i.e., (8) could be satisfied. In addition, the cumulative distribution functions (CDFs) of energy levels in ESSs are provided (note that just the results under \emph{Proposed-2} with $w=10^{-12}$ are given) in Fig.~\ref{fig_3}~(c), where energy levels fluctuate within their normal ranges, i.e., (15) could be guaranteed. Based on the above description, it can be known that the solution of the proposed algorithm is feasible to the original problem \textbf{P1}.

\subsubsection{Convergence results}
Before giving the performance comparisons between the proposed algorithm and other baselines, we first provide the convergence results of the proposed algorithm, which are illustrated in Figs.~\ref{fig_4}~(a)-(c). In Fig.~\ref{fig_4}~(a), the iterative process of the total operational cost in a time slot is shown, while Figs.~\ref{fig_4}~(b) and (c) show the trajectory of the primal residual and feasibility violation metric (which are defined in Appendix C), respectively. It can be observed that the proposed algorithm converges to the same optimal value (which is the same as the result generated by the GAMS commercial solver\footnote{http://www.gams.com/}) given different penalty parameters $\rho$. Moreover, the computation complexity of the proposed algorithm is low since all subproblems in the distributed implementation could be solved in parallel based on closed-form expressions or binary search. Since we do not have enough hardware resources to conduct an experiment with a parallel implementation, the proposed algorithm is implemented on a single Intel Core i5-2410M 2.3GHz server (4G RAM), it takes 1.462 seconds to finish 600 iterations. Since the duration of a time slot is usually several minutes/hours (e.g., electricity prices in some deregulated electricity markets are updated every 5 minutes\footnote{http://www.pjm.com/pub/account/lmpgen/lmppost.html}), the time consumed by the proposed algorithm could be neglected when considering parallel implementation and ``early braking" (i.e., terminating the algorithm before the convergence is reached once we obtain an acceptable solution, e.g., the primal residual and feasibility violation are small enough). Therefore, the proposed online distributed algorithm is very suitable for practical applications.

\subsubsection{Queue weight $w$}
In Fig.~\ref{fig_5}~(a), the operational costs under different algorithms are provided, and we find that the proposed algorithm achieves the best performance. Compared with \emph{B1}, \emph{B2}, and \emph{B3}, \emph{Proposed-2} with $w=10^{-12}$ can reduce the operational cost by 1.48\%, 2.55\%, and 15.15\%, respectively. The reason is that the proposed algorithm can fully utilize the temporal diversity of electricity price by serving batch workloads in proper time slots without violating their deadlines, by controlling the discharging/charging of ESSs in proper time slots, and by selling electricity to main grids when there are excess renewable energies. Thus, the proposed algorithm could obtain the largest profit of selling electricity among all algorithms as shown in Fig.~\ref{fig_5}~(b). In addition, it can be observed that larger $w$ results in smaller AMQD (The Average value of Maximum Queueing Delays experienced by all workloads $\pi_{i,q,t}$), since larger $w$ would lead to more frequent service for batch workloads as indicated in the objective function of \textbf{P5} in Appendix E, which means that less temporal diversity of electricity price could be utilized to reduce operational cost. Consequently, the proposed algorithm shows better performances given a smaller $w$.

\subsubsection{Dropping penalty factor $\theta_i$}
We set $w=10^{-12}$ in this scenario. In Figs.~\ref{fig_6}~(a) and (b), it can be seen that \emph{Proposed-2} always achieves the lowest operational cost. By observing the objective function of \textbf{P6}, it can be known that the proposed algorithm intends to discard less batch workloads given a larger $\theta_i$, resulting in a smaller dropping ratio (i.e., $\sum\nolimits_i\sum\nolimits_q\sum\nolimits_t (e_{i,q,t}/a_{i,q,t})$) as shown in Fig.~\ref{fig_6}~(c). Therefore, the proposed algorithm would reduce to be \emph{B1} if $\theta_i$ is approaching to zero, since all batch workloads would be dropped and no energy queue is needed under this situation.

\subsubsection{Tolerant service delay $\mathcal{T}_{i,q}$}
For simplicity, we assume that $\mathcal{T}_{i,q}$ is the same for all $i$ and $q$. As shown in Figs.~\ref{fig_7}~(a) and (b), the operational cost becomes lower and the profit of selling electricity become larger with the increase of tolerant service delay if $w=10^{-10}$, while those values are almost unchanged if $w=10^{-5}$. The reason is that the proposed algorithm puts very large ``weight" on maintaining the stability of workload queue $Q_{i,q,t}$ and virtual queue $H_{i,q,t}$ when $w=10^{-5}$, resulting in very small queueing delay and AMQD as shown in Fig.~\ref{fig_7} (c). Consequently, low utilization of temporal price diversity is incurred even the tolerant service delays of batch workloads are large. Thus, choosing a proper queue weight $w$ is critical to utilize the heterogeneous tolerant service delays for operational cost reduction.

\section{Conclusions}\label{s6}
 This paper proposed a distributed realtime algorithm for minimizing the long-term operational cost of multiple data center microgrids with the considerations of many factors, e.g., providing heterogeneous service delay guarantees for batch workloads, interactive workload allocation, batch workload shedding, electricity buying/selling, battery charging/discharging efficiency, and the ramping constraints of backup generators. The proposed algorithm does not require any prior knowledge of statistical characteristics related to system parameters and has low computational complexity. Extensive simulation results showed that the proposed algorithm could reduce the operational cost of data center microgrids effectively.

\appendices

\section{Proof of Lemma 1}
\begin{IEEEproof}
Given a slot $t$, it can be proved that the energy demand $\pi_{i,q,t}$ could be satisfied before $t+R_{i,q}^{\max}$. If the above declaration is not true (a contradiction would be reached), we have $Q_{i,q,\tau}>x_{i,q,\tau}$ for all slots $\tau \in \{t+1,t+2,\cdots,t+R_{i,q}^{\max}\}$. According to \eqref{f_14}, we can obtain that $H_{i,q,\tau+1}=[H_{i,q,\tau}-x_{i,q,\tau}+\varepsilon_{i,q}]^+$ for all slots $\tau \in \{t+1,t+2,\cdots,t+R_{i,q}^{\max}\}$. Continually, we have
\begin{align} \label{af_1}
H_{i,q,\tau+1}\geq H_{i,q,\tau}-x_{i,q,\tau}+\varepsilon_{i,q},
\end{align}

Summing the above equation from slot $t+1$ to $t+R_{i,q}^{\max}$, we have
\begin{align} \label{af_2}
H_{i,q,t+R_{i,q}^{\max}+1}-H_{i,q,t+1}\geq R_{i,q}^{\max}\varepsilon_{i,q}-\sum\limits_{\tau=t+1}^{t+R_{i,q}^{\max}}x_{i,q,\tau}.
\end{align}

Since $H_{i,q,t+1}\geq 0$ and $H_{i,q,t+R_{i,q}^{\max}+1}\leq H_{i,q}^{\max}$, \eqref{af_2} could be transformed into \eqref{af_3},
\begin{align} \label{af_3}
\sum\limits_{\tau=t+1}^{t+R_{i,q}^{\max}}x_{i,q,\tau}+H_{i,q}^{\max}\geq R_{i,q}^{\max}\varepsilon_{i,q}.
\end{align}

In addition, the summation of $x_{i,q,\tau}$ over the interval $\{t+1,t+2,\cdots,t+R_{i,q}^{\max}\}$ is strictly smaller than $Q_{i,q,t+1}$. Otherwise, $\pi_{i,q,t}$ would be served within the interval. Thus, we have
\begin{align} \label{af_4}
\sum\limits_{\tau=t+1}^{t+R_{i,q}^{\max}}x_{i,q,\tau}<Q_{i,q,t+1}\leq Q_{i,q}^{\max}.
\end{align}
Finally, combining \eqref{af_3} and \eqref{af_4}, we obtain
\begin{align} \label{af_5}
R_{i,q}^{\max}<\lceil(H_{i,q}^{\max}+Q_{i,q}^{\max})/\varepsilon_{i,q}\rceil.
\end{align}
Note that \eqref{af_5} contradicts the definition of $R_{i,q}^{\max}$. Thus, the workload batch $\pi_{i,q,t}$ must be served before $t+R_{i,q}^{\max}$.
\end{IEEEproof}

\section{Proof of Lemma 2}
\begin{IEEEproof}
According to the definition of $Q_{i,q,t}$, we have
\begin{align}
Q_{i,q,t+1}^2&=\Big(\max\{Q_{i,q,t}-x_{i,q,t},0\}+\pi_{i,q,t}\Big)^2 \nonumber \\
&\leq Q_{i,q,t}^2+x_{i,q,t}^2+\pi_{i,q,t}^2+2Q_{i,q,t}(\pi_{i,q,t}-x_{i,q,t}). \nonumber
\end{align}

Then, we can obtain
\begin{align}
\frac{Q_{i,q,t+1}^2-Q_{i,q,t}^2}{2}\leq &\frac{(x_{i,q}^{\max})^2+(\pi_{i,q}^{\max})^2}{2}\nonumber \\
&+Q_{i,q,t}(\pi_{i,q,t}-x_{i,q,t}). \nonumber
\end{align}

For the queue $H_{i,q,t}$, we have
\begin{align}
H_{i,q,t+1}^2\leq &\Big(\max[H_{i,q,t}-x_{i,q,t}+\varepsilon_{i,q},0]\Big)^2 \nonumber \\
\leq &\Big(H_{i,q,t}-x_{i,q,t}+\varepsilon_{i,q}\Big)^2. \nonumber
\end{align}

Then, we have
\begin{align}
\frac{H_{i,q,t+1}^2-H_{i,q,t}^2}{2}\leq &\frac{(\varepsilon_{i,q}-x_{i,q,t})^2}{2}+H_{i,q,t}(\varepsilon_{i,q}-x_{i,q,t}), \nonumber\\
\leq &\frac{\max\{\varepsilon_{i,q}^2,(x_{i,q}^{\max})^2\}}{2}+H_{i,q,t}(\varepsilon_{i,q}-x_{i,q,t}). \nonumber
\end{align}

Similarly, for the queue $Z_{i,t}$, we have
\begin{align}
\frac{Z_{i,t+1}^2-Z_{i,t}^2}{2}\leq &\frac{\max\{(\eta_{c,i} u_{i,\text{cmax}})^2,(\frac{1}{\eta_{d,i}}u_{i,\text{dmax}})^2\}}{2}\nonumber\\
&+Z_{i,t}(\eta_{c,i} u_{c,i,t}-\frac{1}{\eta_{d,i}}u_{d,i,t}). \nonumber
\end{align}

Combining three upper bounds mentioned above together, we have the following inequality,
\begin{align}
\Delta_t\leq &\mathbb{E}\{\sum\limits_{i=1}^{N}\sum\limits_{q=1}^{M_i}wQ_{i,q,t}(\pi_{i,q,t}-x_{i,q,t})|\boldsymbol{\Theta_t}\} \nonumber \\
&+\mathbb{E}\{\sum\limits_{i=1}^{N}\sum\limits_{q=1}^{M_i}wH_{i,q,t}(\varepsilon_{i,q}-x_{i,q,t})|\boldsymbol{\Theta_t}\} \nonumber \\
&+\mathbb{E}\{\sum\limits_{i=1}^{N}Z_{i,t}(\eta_{c,i} u_{c,i,t}-\frac{1}{\eta_{d,i}}u_{d,i,t})|\boldsymbol{\Theta_t}\}+\Omega_0.
\end{align}
By adding $V\mathbb{E}\{\Gamma_{t}|\boldsymbol{\Theta_t}\}$ to the both sides of the above equation, we could complete the proof.
\end{IEEEproof}

\section{The distributed implementation of Algorithm 1}
\begin{IEEEproof}
\underline{\textbf{1. Initialization}}: Decision variables of \textbf{P6} are initialized with zero. In each iteration $k$, two steps (i.e., prediction step and correction step) are repeated until convergence.

\underline{\textbf{2. ADMM step (prediction step)}}. Obtain all decision variables in the forwarding order:

\underline{\textbf{2.1 $d_{f,i,t}$-minimization}}: each front-end server $f$ solves \textbf{P7} in parallel to obtain $\tilde{d}_{f,i,t}^k$.
  \begin{subequations}\label{acf_1}
\begin{align}
(\textbf{P7})~\min~~&\Phi_1(d_{f,i,t},\chi_{f,i}^k,a_{f,i,t}^k) \\
s.t.~&(1),(2),
\end{align}
\end{subequations}
where $\Phi_1(d_{f,i,t},\chi_{i,q}^k,a_{f,i,t}^k)=\sum\nolimits_{i=1}^N\big((V\omega L_{f,i}+\chi_{f,i}^k-\rho a_{f,i,t}^k)d_{f,i,t}+\frac{\rho}{2}d_{f,i,t}^2\big)$; $\rho$ is the penalty parameter in the augmented Lagrangian for \textbf{P6}, while $\phi_i$,$\varphi_i$,$\kappa_{i,q}$,$\chi_{f,i}$,$\psi_{i,q}$ are dual variables associated with (37c,-(37g), respectively.

\underline{\textbf{2.2 $x_{i,q,t}$-minimization}}: each queue controller $q$ in data center $i$ solves \textbf{P8} in parallel to obtain $\tilde{x}_{i,q,t}^k$.
  \begin{subequations}\label{acf_2}
\begin{align}
(\textbf{P8})~\min~~&\Phi_2(x_{i,q,t},\psi_{i,q}^k,b_{i,q,t}^k)\\
s.t.~&(4),
\end{align}
\end{subequations}
where $\Phi_2(x_{i,q,t},\psi_{i,q}^k,b_{i,q,t}^k)=\psi_{i,q}^kx_{i,q,t}+\frac{\rho}{2}(x_{i,q,t}-b_{i,q,t}^k)^2$.

\underline{\textbf{2.3 $c_{i,t}$-minimization}}: each conventional generator in SMG $i$ solves \textbf{P9} in parallel to obtain $\tilde{c}_{i,t}^k$.
  \begin{subequations}\label{acf_3}
\begin{align}
(\textbf{P9})~\min~~&\Phi_3(c_{i,t},\varphi_i^k,g_{i,t}^k,u_{d,i,t}^k,u_{c,i,t}^k,h_{i}^k) \\
s.t.~&(10),(11),
\end{align}
\end{subequations}
where $\Phi_3(c_{i,t},\varphi_i^k,g_{i,t}^k,u_{d,i,t}^k,u_{c,i,t}^k,h_{i}^k)=VA_i(c_{i,t})+\frac{\rho}{2}c_{i,t}^2+(\varphi_i^k+\rho(g_{i,t}^k+u_{d,i,t}^k-u_{c,i,t}^k+\beta_ih_{i}^k-m_i))c_{i,t}$.

\underline{\textbf{2.4 $u_{c,i,t}$-minimization}}: each ESS in SMG $i$ solves \textbf{P10} in parallel to obtain $\tilde{u}_{c,i,t}^k$.
  \begin{subequations}\label{acf_4}
\begin{align}
(\textbf{P10})~&\min~~\Phi_4(u_{c,i,t},\varphi_i^k,u_{d,i,t}^k,g_{i,t}^k,\tilde{c}_{i,t}^k,h_i^k)\\
s.t.&~(12),
\end{align}
\end{subequations}
where $\Phi_4(u_{c,i,t},\varphi_i^k,u_{d,i,t}^k,g_{i,t}^k,\tilde{c}_{i,t}^k,h_i^k)=\frac{\rho}{2}u_{c,i,t}^2 +VB_i(u_{c,i,t},u_{d,i,t}^k)+(Z_{i,t}\eta_{c,i}-\varphi_i^k-\rho(g_{i,t}^k+\tilde{c}_{i,t}^k+u_{d,i,t}^k+\beta_ih_{i}^k-m_i))u_{c,i,t}$.

\underline{\textbf{2.5 $u_{d,i,t}$-minimization}}: each ESS in SMG $i$ solves \textbf{P11} in parallel to obtain $\tilde{u}_{d,i,t}^k$.
\begin{subequations}\label{acf_5}
\begin{align}
(\textbf{P11})~&\min~~\Phi_5(u_{d,i,t},\tilde{u}_{c,i,t}^k,\varphi_i^k,g_{i,t}^k,\tilde{c}_{i,t}^k,h_i^k)\\
s.t.&~(13),
\end{align}
\end{subequations}
where $\Phi_5(u_{d,i,t},\tilde{u}_{c,i,t}^k,\varphi_i^k,g_{i,t}^k,\tilde{c}_{i,t}^k,h_i^k)=\frac{\rho}{2}u_{d,i,t}^2 +VB_i(\tilde{u}_{c,i,t}^k,u_{d,i,t})-(Z_{i,t}/\eta_{d,i}-\varphi_i^k-\rho(g_{i,t}^k+\tilde{c}_{i,t}^k-\tilde{u}_{c,i,t}^k+\beta_ih_{i}^k-m_i))u_{d,i,t}$.

\underline{\textbf{2.6 $a_{f,i,t}$-minimization}}: each EMS in SMG $i$ solves \textbf{P12} in parallel to obtain $\tilde{a}_{f,i,t}^k$.
\begin{subequations}\label{acf_6}
\begin{align}
(\textbf{P12})~&\min~~\Phi_6(a_{f,i,t},\phi_i^k,b_{i,q,t}^k,e_{i,q,t}^k,h_i^k,\chi_{f,i}^k,\tilde{d}_{f,i,t}^k)\\
s.t.&~a_{f,i,t}\geq 0,
\end{align}
\end{subequations}
where $\Phi_6(a_{f,i,t},\phi_i^k,b_{i,q,t}^k,e_{i,q,t}^k,h_i^k,\chi_{f,i}^k,\tilde{d}_{f,i,t}^k)=\frac{\rho}{2}(\sum\nolimits_{f=1}^F a_{f,i,t}^2+(\sum\nolimits_{f=1}^F a_{f,i,t})^2)+\sum\nolimits_{f=1}^F(\phi_i^k-\chi_{f,i}^k-\rho \tilde{d}_{f,i,t}^k+\rho(\sum\limits_{q=1}^{M_i}(b_{i,q,t}^k-e_{i,q,t}^k)+h_i^k-C_i)a_{f,i,t}$.

\underline{\textbf{2.7 $b_{i,q,t}$-minimization}}: each EMS in SMG $i$ solves \textbf{P13} in parallel to obtain $\tilde{b}_{i,q,t}^k$.
\begin{subequations}\label{acf_7}
\begin{align}
(\textbf{P13})~&\min~~\Phi_7(b_{i,q,t},\phi_i^k,\kappa_{i,q}^k,\psi_{i,q}^k,\tilde{a}_{f,i,t}^k,e_{i,q,t}^k,h_i^k,z_{i,q}^k)\\
s.t.&~b_{i,q,t}\geq 0,
\end{align}
\end{subequations}
where $\Phi_7(b_{i,q,t},\phi_i^k,\kappa_{i,q}^k,\tilde{a}_{f,i,t}^k,e_{i,q,t}^k,h_i^k,z_{i,q}^k)=\frac{\rho}{2}(\sum\nolimits_{q=1}^{M_i} 2b_{i,q,t}^2+(\sum\nolimits_{q=1}^{M_i} b_{i,q,t})^2)-\sum\nolimits_{q=1}^{M_i}(w(Q_{i,q,t}+H_{i,q,t})-\phi_i^k+\kappa_{i,q}^k+\psi_{i,q}^k+\rho(e_{i,q,t}^k+z_{i,q}^k+\tilde{x}_{i,q,t}^k)-\rho(\sum\nolimits_{f=1}^F \tilde{a}_{f,i,t}^k-\sum\nolimits_{q=1}^{M_i}e_{i,q,t}^k+h_i^k-C_i))b_{i,q,t}$.

\underline{\textbf{2.8 $e_{i,q,t}$-minimization}}: each EMS in SMG $i$ solves \textbf{P14} in parallel to obtain $\tilde{e}_{i,q,t}^k$.
\begin{subequations}\label{acf_8}
\begin{align}
(\textbf{P14})~&\min~~\Phi_8(e_{i,q,t},\kappa_{i,q}^k,\phi_i^k,z_{i,q}^k,\tilde{b}_{i,q,t}^k,\tilde{a}_{f,i,t}^k,h_i^k)\\
s.t.&~e_{i,q,t}\geq 0,
\end{align}
\end{subequations}
where $\Phi_8(e_{i,q,t},\kappa_{i,q}^k,\phi_i^k,z_{i,q}^k,\tilde{b}_{i,q,t}^k,\tilde{a}_{f,i,t}^k,h_i^k)=\frac{\rho}{2}(\sum\nolimits_{q=1}^{M_i} e_{i,q,t}^2+(\sum\nolimits_{q=1}^{M_i} e_{i,q,t})^2)+\sum\nolimits_{q=1}^{M_i}(V\theta_i-\phi_i^k+\kappa_{i,q}^k+\rho(z_{i,q}^k-\tilde{b}_{i,q,t}^k)-\rho(\sum\nolimits_{f=1}^F \tilde{a}_{f,i,t}^k+\sum\nolimits_{q=1}^{M_i}\tilde{b}_{i,q,t}^k+h_i^k-C_i))e_{i,q,t})$.

\underline{\textbf{2.9 $h_i$-minimization}}: each EMS in SMG $i$ solves \textbf{P15} in parallel to obtain $\tilde{h}_{i}^k$.
\begin{subequations}\label{acf_9}
\begin{align}
(\textbf{P15})~&\min~\Phi_9(h_i,\phi_i^k,\varphi_i^k,g_{i,t}^k,\tilde{c}_{i,t}^k,\tilde{u}_{c,i,t}^k,\tilde{u}_{d,i,t}^k,\tilde{a}_{f,i,t}^k,\tilde{b}_{i,q,t}^k,\tilde{e}_{i,q,t}^k)\\
s.t.&~h_i\geq 0,
\end{align}
\end{subequations}
where $\Phi_9(h_i,\phi_i^k,\varphi_i^k,g_{i,t}^k,\tilde{c}_{i,t}^k,\tilde{u}_{c,i,t}^k,\tilde{u}_{d,i,t}^k,\tilde{a}_{f,i,t}^k,\tilde{b}_{i,q,t}^k,\tilde{e}_{i,q,t}^k)=\frac{\rho}{2}(1+\beta_i^2)h_i^2+(\phi_i^k+\beta_i\varphi_i^k+\rho\beta_i(g_{i,t}^k+\tilde{c}_{i,t}^k+\tilde{u}_{d,i,t}^k-\tilde{u}_{c,i,t}^k-m_i)+\rho(\sum\nolimits_{f=1}^F \tilde{a}_{f,i,t}^k+\sum\nolimits_{q=1}^{M_i}(\tilde{b}_{i,q,t}^k-\tilde{e}_{i,q,t}^k)-C_i))h_i$.

\underline{\textbf{2.10 $z_{i,q}$-minimization}}: each EMS in SMG $i$ solves \textbf{P16} in parallel to obtain $\tilde{z}_{i,q}^k$.
\begin{subequations}\label{acf_10}
\begin{align}
(\textbf{P16})~&\min~~\Phi_{10}(z_{i,q},\tilde{e}_{i,q,t}^k,\tilde{b}_{i,q,t}^k,\kappa_{i,q}^k)\\
s.t.&~z_{i,q}\geq 0,
\end{align}
\end{subequations}
where $\Phi_{10}(z_{i,q},\tilde{e}_{i,q,t}^k,\tilde{b}_{i,q,t}^k,\kappa_{i,q}^k)=\frac{\rho}{2}z_{i,q}^2+(\rho(\tilde{e}_{i,q,t}^k-\tilde{b}_{i,q,t}^k)+\kappa_{i,q}^k)z_{i,q}$.

\underline{\textbf{2.11 $g_{i,t}$-minimization}}: each EMS in SMG $i$ solves \textbf{P17} in parallel to obtain $\tilde{g}_{i,t}^k$.
\begin{subequations}\label{acf_11}
\begin{align}
(\textbf{P17})~&\min~~\Phi_{11}(g_{i,t},\varphi_i^k,\tilde{c}_{i,t}^k,\tilde{u}_{c,i,t}^k,\tilde{u}_{d,i,t}^k,\tilde{h}_{i}^k)\\
s.t.&~(18),
\end{align}
\end{subequations}
where $\Phi_{11}(g_{i,t},\tilde{u}_{c,i,t}^k,\tilde{u}_{d,i,t}^k,\tilde{h}_{i}^k)=\frac{\rho}{2}g_{i,t}^2+(\varphi_i^k+\rho(\tilde{c}_{i,t}^k+\tilde{u}_{d,i,t}^k-\tilde{u}_{c,i,t}^k+\beta_i\tilde{h}_{i}^k-m_i))g_{i,t}+\frac{X_{i,t}-W_{i,t}}{2}|g_{i,t}|+\frac{X_{i,t}+W_{i,t}}{2}g_{i,t}$.

Note that \textbf{P7}-\textbf{P17} are convex optimization problems and their solutions could be obtained easily based on closed-form expressions or binary search. Thus, the algorithms for them are omitted for brevity. Similar algorithms could be found in \cite{Liang2016TSG}.

\underline{\textbf{2.12 Dual update}}: the EMS in SMG $i$ updates $\tilde{\phi}_i^k$,$\tilde{\varphi}_{i}^k$,$\tilde{\kappa}_{i,q}^k$ as follows: $\tilde{\phi}_i^k=\phi_i^k+\rho(\sum\nolimits_{f=1}^F \tilde{a}_{f,i,t}+\sum\nolimits_{q=1}^{M_i}(\tilde{b}_{i,q,t}-\tilde{e}_{i,q,t})+\tilde{h}_i-C_i)$; $\tilde{\varphi}_{i}^k=\varphi_{i}^k+\rho(\tilde{g}_{i,t}^k+\tilde{c}_{i,t}^k+\tilde{u}_{d,i,t}^k-\tilde{u}_{c,i,t}^k+\beta_i \tilde{h}_i^k-m_i)$; $\tilde{\kappa}_{i,q}^k=\kappa_{i,q}^k+\rho(\tilde{e}_{i,q,t}^k+\tilde{z}_{i,q}^k-\tilde{b}_{i,q,t}^k)$; each front-end server $f$ updates $\tilde{\chi}_{f,i}^k$ as follows, i.e., $\tilde{\chi}_{f,i}^k=\chi_{f,i}^k+\rho(\tilde{d}_{f,i,t}^k-\tilde{a}_{f,i,t}^k)$; each queue controller $q$ in data center $i$ updates $\tilde{\psi}_{i,q}^k$ as follows: $\tilde{\psi}_{i,q}^k=\psi_{i,q}^k+\rho(\tilde{x}_{i,q,t}^k-\tilde{b}_{i,q,t}^k)$.

\underline{\textbf{3. Gaussian back substitution step (correction step)}}:
Obtain the input parameters of iteration $k+1$ according to the Gaussian back substitution step (3.5b) in \cite{He2012}, where the constant $\alpha$ in (3.5b) is set to one based on the practical experience\cite{Liang2016TSG}. Then, we have
\begin{align}
&\phi_i^{k+1}=\tilde{\phi}_i^k,\varphi_i^{k+1}=\tilde{\varphi}_{i}^k,\kappa_{i,q}^{k+1}=\tilde{\kappa}_{i,q}^k,\nonumber \\
&\tilde{\chi}_{f,i}^k=\chi_{f,i}^k,~\psi_{i,q}^{k+1}=\tilde{\psi}_{i,q}^k,g_{i,t}^{k+1}=\tilde{g}_{i,t}^k, z_{i,q}^{k+1}=\tilde{z}_{i,q}^k, \nonumber \\
&h_i^{k+1}=\tilde{h}_i^k-\frac{\beta_i}{1+\beta_i^2}(\tilde{g}_{i,t}^{k}-g_{i,t}^{k}),\nonumber \\
&e_{i,q,t}^{k+1}=\tilde{e}_{i,q,t}^k+\frac{(h_i^{k+1}-h_i^k)+\sum\nolimits_{q=1}^{M_i}(\tilde{z}_{i,q}^{k}-z_{i,q}^k)}{M_i+1}-(\tilde{z}_{i,q}^{k}-z_{i,q}^k), \nonumber \\
&b_{i,q,t}^{k+1}=\tilde{b}_{i,q,t}^k+\frac{\sum\limits_{q=1}^{M_i}(e_{i,q,t}^{k+1}-e_{i,q,t}^k-z_{i,q}^{k+1}+z_{i,q}^k)-2(h_i^{k+1}-h_i^k)}{2(M_i+2)}\nonumber \\
&~~~~~~~~+\frac{1}{2}(z_{i,q}^{k+1}-z_{i,q}^{k}+e_{i,q,t}^{k+1}-e_{i,q,t}^{k}),\nonumber \\
&a_{f,i,t}^{k+1}=\tilde{a}_{f,i,t}^{k}+\frac{\sum\limits_{q=1}^{M_i}(e_{i,q,t}^{k+1}-e_{i,q,t}^{k}-b_{i,q,t}^{k+1}+b_{i,q,t}^{k})-(h_i^{k+1}-h_i^k)}{F+1}\nonumber \\
&u_{d,i,t}^{k+1}=\tilde{u}_{d,i,t}^{k}-\beta_i(h_i^{k+1}-h_i^k)-(g_{i,t}^{k+1}-g_{i,t}^k),\nonumber \\
&u_{c,i,t}^{k+1}=\tilde{u}_{c,i,t}^{k}+(\tilde{u}_{d,i,t}^{k}-u_{d,i,t}^{k}),~c_{i,t}^{k+1}=\tilde{c}_{i,t}^k+(\tilde{u}_{c,i,t}^{k}-u_{c,i,t}^{k}),\nonumber \\
&x_{i,q,t}^{k+1}=\tilde{x}_{i,q,t}^k+(b_{i,q,t}^{k+1}-b_{i,q,t}^{k}),~d_{f,i,t}^{k+1}=\tilde{d}_{f,i,t}^{k}.\nonumber
\end{align}

\underline{\textbf{4. Stopping criterion}}: As in our previous work\cite{Liang2016TSG}, we terminate the designed algorithm before the convergence is reached once we obtain an acceptable feasible solution, e.g., when the primal residual $\Xi$ is small enough and the obtained solution is feasible. Specifically, the primal residual is defined in \eqref{acf_12}. Moreover, a feasibility metric $\ell$ is adopted as in \eqref{acf_13} to indicate the feasibility of the obtained solution.

\begin{align}\label{acf_12}
\Xi^2=&\sum\limits_{i=1}^{N}\Big(\sum\limits_{f=1}^F a_{f,i,t}^k+\sum\limits_{q=1}^{M_i}(b_{i,q,t}^k-e_{i,q,t}^k)+h_i^k-C_i\Big)^2 \nonumber\\
+&\sum\limits_{i=1}^{N}\Big(g_{i,t}^k+c_{i,t}^k+u_{d,i,t}^k-u_{c,i,t}^k+\beta_i h_i^k-m_i\Big)^2 \nonumber\\
+&\sum\limits_{i=1}^{N}\sum\limits_{q=1}^{M_i}\Big(e_{i,q,t}^k+z_{i,q}^k-b_{i,q,t}^k \Big)^2\nonumber\\
+&\sum\limits_{i=1}^{N}\sum\limits_{q=1}^{M_i}\Big(x_{i,q,t}^k-b_{i,q,t}^k \Big)^2\nonumber\\
+&\sum\limits_{i=1}^{N}\sum\limits_{f=1}^{F}\Big(d_{f,i,t}^k-a_{f,i,t}^k \Big)^2.
 \end{align}

\begin{align} \label{acf_13}
\ell=&\sum\limits_{i = 1}^N \max\Big(\sum\limits_{f=1}^F d_{f,i,t}^k+\sum\limits_{q=1}^{M_i}(x_{i,q,t}^k-e_{i,q,t}^k)-C_i,0\Big)\nonumber \\
&+\sum\limits_{i = 1}^N \Big|g_{i,t}^k+r_{i,t}+c_{i,t}^k+u_{d,i,t}^k-p_{i,t}^k-u_{c,i,t}^k\Big|\nonumber \\
&+\sum\limits_{i = 1}^N\sum\limits_{q=1}^{M_i}\max\Big(e_{i,q,t}^k-x_{i,q,t}^k,0\Big).
\end{align}

Note that the implementation of one iteration in the proposed algorithm could be described as follows. At initial iteration, each front-end server $f$, each queue controller $q$ in data center $i$, each conventional generator in SMG $i$ make their local and parallel decisions independently to obtain $\tilde{a}_{f,i,t}^k$, $\tilde{x}_{i,q,t}^k$, $\tilde{c}_{i,t}^k$, respectively. Then, such decisions are broadcasted to other components in SMGs, e.g., ESSs and EMS. After receiving the broadcasted decisions, each ESS and EMS in SMG $i$ make their local decisions on $\tilde{a}_{f,i,t}^{k}$ and $\tilde{u}_{c,i,t}^{k}$, and $\tilde{u}_{d,i,t}^{k}$. Then, ESS $i$ broadcasts $\tilde{u}_{c,i,t}^{k}$, and $\tilde{u}_{d,i,t}^{k}$ to the EMS $i$. Next, EMS $i$ could obtains other decisions on $\tilde{b}_{i,q,t}^{k}$, $\tilde{e}_{i,q,t}^{k}$, $\tilde{h}_i^{k}$, $\tilde{z}_{i,q}^k$ and $\tilde{g}_{i,t}^k$. Finally, EMS $i$ broadcasts all obtained decision variables at iteration $k+1$ (i.e., $g_{i,t}^{k+1}$, $z_{i,q}^{k+1}$, $h_i^{k+1}$, $e_{i,q,t}^{k+1}$, $b_{i,q,t}^{k+1}$, $a_{f,i,t}^{k+1}$, $u_{d,i,t}^{k+1}$,$u_{c,i,t}^{k+1}$) so that all entities (i.e., front-end servers, queue controllers, conventional generators, and ESSs) could update their respective decisions in iteration $k+1$ according to the Gaussian back substitution step.

\end{IEEEproof}

\section{Proof of Lemma 3}
\begin{IEEEproof}
\begin{enumerate}
 \item Let $(u_{c,i,t}^{*},u_{d,i,t}^{*},x_{i,q,t}^*,e_{i,q,t}^*,d_{f,i,t}^*,\pi_{i,t}^*,g_{i,t}^*,c_{i,t}^*)$ be the optimal decision vector obtained from the Algorithm 1. For SMG $i$, suppose $Z_{i,t}<-V\eta_{d,i}\gamma_{i,\max}$ and $u_{d,i,t}^{*}>0$, then, $u_{c,i,t}^{*}=0$.
     Then, we can prove the non-optimality of the above decision by choosing another decision vector $(0,0,x_{i,q,t}^*,e_{i,q,t}^*,d_{f,i,t}^*,\pi_{i,t}^*,\tilde{g}_{i,t}^*,\tilde{c}_{i,t}^*)$. Suppose the objective values corresponding to the above decision vectors under the Algorithm 1 are $\Upsilon_{1,i}$ and $\Upsilon_{2,i}$, respectively. Given the same energy demand $p_{i,t}$, there are three kinds of the decisions for energy supply:  \\
    \textbf{Case 1}: If $g_{i,t}^*=0$, we choose $\tilde{g}_{i,t}^*=0$, then,
    $\tilde{c}_{i,t}^*=c_{i,t}^*+u_{d,i,t}^{*}$. Next, $\Upsilon_{1,i}-\Upsilon_{2,i}>  (-\frac{Z_{i,t}}{\eta_{d,i}}-VA_{i,\max}')u_{d,i,t}^{*}>0$.\\
     \textbf{Case 2}: If $g_{i,t}^*>0$, we choose $\tilde{g}_{i,t}^*=0$, $\tilde{c}_{i,t}^*=c_{i,t}^*$, then, $\tilde{g}_{i,t}^*=g_{i,t}^*+u_{d,i,t}^{*}$. Next, $\Upsilon_{1,i}-\Upsilon_{2,i}> (-\frac{Z_{i,t}}{\eta_{d,i}}-VX_{i,\max})u_{d,i,t}^{*}>0$.\\
    \textbf{Case 3}: If $g_{i,t}^*<0$, we choose $\tilde{g}_{i,t}^*=0$, $\tilde{c}_{i,t}^*=c_{i,t}^*$, then, $\tilde{g}_{i,t}^*=g_{i,t}^*-u_{d,i,t}^{*}$. Next, $\Upsilon_{1,i}-\Upsilon_{2,i}> (-\frac{Z_{i,t}}{\eta_{d,i}}-VW_{i,\max})u_{d,i,t}^{*}>0$.\\
     In summary, when $Z_{i,t}<-V\eta_{d,i}\gamma_{i,\max}$, the optimal discharging decision is $u_{d,i,t}^{*}=0$.
 \item The proof of part 2 is similar to that of part 1. Thus, it is omitted for brevity.
\end{enumerate}
\end{IEEEproof}

\section{Proof of Theorem 1}
\begin{IEEEproof}
\begin{enumerate}
  \item The objective value of \textbf{P5} could be rewritten as follows by discarding some constant items,
        \begin{align}
        &\bigsqcup(d_{f,i,t}, e_{i,q,t}, c_{i,t}, g_{i,t}, u_{c,i,t}, u_{d,i,t})\nonumber \\
        &+\sum\limits_{i=1}^{N}\sum\limits_{q=1}^{M_i}(V\beta_iX_{i,t}-w(Q_{i,q,t}+H_{i,q,t}))x_{i,q,t}\mathcal{I}_1,\nonumber \\
        &+\sum\limits_{i=1}^{N}\sum\limits_{q=1}^{M_i}(V\beta_iW_{i,t}-w(Q_{i,q,t}+H_{i,q,t}))x_{i,q,t}\mathcal{I}_2,\nonumber \\
        &+\sum\limits_{i=1}^{N}\sum\limits_{q=1}^{M_i}(-w(Q_{i,q,t}+H_{i,q,t}))x_{i,q,t}\mathcal{I}_3,\nonumber
      \end{align}
      where $\bigsqcup(\upsilon)$ is the function of $\upsilon$; $\mathcal{I}_1$, $\mathcal{I}_2$, $\mathcal{I}_3$ denote $g_{i,t}>0$, $g_{i,t}<0$, and $g_{i,t}=0$, respectively. It can be observed that the proposed algorithm would choose the maximum possible $x_{i,q,t}$ when $Q_{i,q,t}>V\beta_iX_i^{\max}/w$.
      In the following parts, we would use the induction method to prove $Q_{i,q}^{\max}=V\beta_iX_i^{\max}/w+\pi_{i,q}^{\max}$ for all slots. It is obvious that $Q_{i,q,0}\leq Q_{i,q}^{\max}$. Suppose $Q_{i,q,t}\leq Q_{i,q}^{\max}$, we will show that $Q_{i,q,t+1}\leq Q_{i,q}^{\max}$. If $Q_{i,q,t}\leq V\beta_iX_{i}^{\max}/w$, the maximum queue growth is $\pi_{i,q}^{\max}$. Thus, we have $Q_{i,q,t+1}\leq Q_{i,q,t}+\pi_{i,q}^{\max}\leq V\beta_iX_{i}^{\max}/w+\pi_{i,q}^{\max}$. If $Q_{i,q,t}\geq V\beta_iX_{i}^{\max}/w$, the proposed algorithm would choose $x_{i,q,t}=\min\{Q_{i,q,t},x_{i,q}^{\max}\}$. Thus, $Q_{i,q,t+1}\leq \max\{Q_{i,q,t},\pi_{i,q}^{\max}\}\leq Q_{i,q}^{\max}$. Similarly, we can prove that $Z_{i,q,t}\leq Z_{i,q}^{\max}$ for any slot $t$. The proof detail is omitted for brevity. Continually, it can be known that \eqref{f_5} could be satisfied.
  \item According to Lemma 1 and the part 1 of Theorem 1, we have $R_{i,q}^{\max}=\lceil(2V\beta_iX_{i,q}^{\max}/w+\pi_{i,q}^{\max}+\varepsilon_{i,q})/\varepsilon_{i,q}\rceil$. Therefore, we can construct an algorithm to ensure that all charging requests have delay less than or equal to $R_{i,q}^{\max}$ slots, where $R_{i,q}^{\max}\geq 2$. When choosing $x_{i,q,t}=x_{i,q}^{\max}$ in each time slot $t$, we can guarantee that all charging requests have one slot delay. In summary, the proposed algorithm could be constructed to ensure the heterogeneous service delays for all EV charging requests, i.e., \eqref{f_6} could be satisfied under the proposed algorithm.
  \item Proving $D_{i,t}\in [D_{i,\min},~D_{i,\max}]$ is equivalent to satisfying the following constraints:
         $Z_{i,t}\geq-V\eta_{d,i}\gamma_{i,\max}-\frac{1}{\eta_{d,i}}u_{i,\text{dmax}}$, and $Z_{i,t}\leq D_{i,\max}-D_{i,\min}-V\eta_{d,i}\gamma_{i,\max}-\frac{1}{\eta_{d,i}}u_{i,\text{dmax}}$.
  Because $D_{i,\min}\leq D_{i,0}\leq D_{i,\max}$, the above inequalities hold for $t$=0. Suppose the above-mentioned inequalities hold for the time slot $t$, we should verify that they hold for the time slot $t$+1.
  \begin{itemize}
    \item If $-V\eta_{d,i}\gamma_{i,\max}- \frac{1}{\eta_{d,i}}u_{i,\text{dmax}}\leq Z_{i,t}<-V\eta_{d,i}\gamma_{i,\max}$, then, according to the \text{Lemma 3}, $u_{d,i,t}^{*}=0$. As a result, $Z_{i,t+1}=Z_{i,t}+\eta_{c,i} u_{c,i,t}^{*}\geq Z_{i,t}\geq -V\eta_{d,i}\gamma_{i,\max}- \frac{1}{\eta_{d,i}}u_{i,\text{dmax}}$. If $-V\eta_{d,i}\gamma_{i,\max}\leq Z_{i,t}<D_{i,\max}-D_{i,\min}-V\eta_{d,i}\gamma_{i,\max}- \frac{1}{\eta_{d,i}}u_{i,\text{dmax}}$, then, $Z_{i,t+1} \geq -V\eta_{d,i}\gamma_{i,\max}-\frac{1}{\eta_{d,i}} u_{d,i,t}^{*}>-V\eta_{d,i}\gamma_{i,\max}- \frac{1}{\eta_{d,i}}u_{i,\text{dmax}}$.
    \item If $-\frac{V}{\eta_{d,i}}\gamma_{i,\min}<Z_{i,t}\leq D_{i,\max}-D_{i,\min} - V\eta_{d,i}\gamma_{i,\max}- \frac{1}{\eta_{d,i}}u_{i,\text{dmax}}$, then, $u_{c,i,t}^{*}=0$. Consequently, $Z_{i,t+1}\leq Z_{i,t}\leq D_{i,\max}-D_{i,\min} - V\eta_{d,i}\gamma_{i,\max}- \frac{1}{\eta_{d,i}}u_{i,\text{dmax}}$. If $- V\eta_{d,i}\gamma_{i,\max}- \frac{1}{\eta_{d,i}}u_{i,\text{dmax}}\leq Z_{i,t}\leq -\frac{V}{\eta_{c,i}}\gamma_{i,\min}$, then, $Z_{i,t+1}\leq -\frac{V}{\eta_{c,i}}\gamma_{i,\min}+ \eta_{c,i}u_{i,\text{cmax}}\leq D_{i,\max}-D_{i,\min} - V\eta_{d,i}\gamma_{i,\max}- \frac{1}{\eta_{d,i}}u_{i,\text{dmax}}$, where
    \begin{eqnarray}\label{af_7}
    V\leq \frac{{D_{i,\max}-D_{i,\min} -  (\eta_{c,i}u_{i,\text{cmax}} + \frac{1}{\eta_{d,i}} u_{i,\text{dmax}})}}{\eta_{d,i}\gamma_{i,\max}-\frac{1}{\eta_{c,i}}\gamma_{i,\min}}.\nonumber
    \end{eqnarray}
  \end{itemize}
  Continually, $V_{\max}$ is obtained as follows,
\begin{eqnarray}\label{af_6}
&V_{\max}  = \mathop {\min }\limits_i \frac{{D_{i,\max}-D_{i,\min}  -  (\eta_{c,i}u_{i,\text{cmax}} + \frac{1}{\eta_{d,i}} u_{i,\text{dmax}})}}{\eta_{d,i}\gamma_{i,\max}-\frac{1}{\eta_{c,i}}\gamma_{i,\min}}.\nonumber
\end{eqnarray}
Based on the above proof, it can be known that \eqref{f_12} could be satisfied.
\item From the parts 1-3, we know that the constraints \eqref{f_5},\eqref{f_6},\eqref{f_12} could be satisfied under the proposed algorithm. Since other constraints in \textbf{P1} could be guaranteed, the solution of the proposed algorithm is feasible to the original problem \textbf{P1}.
\item Let ($x_{i,q,t}^{*}$, $u_{c,i,t}^{*},u_{d,i,t}^{*},d_{f,i,t}^{*},\pi_{i,t}^{*},g_{i,t}^{*},c_{i,t}^{*}$) and ($\bar{x}_{i,q,t}$, $\bar{u}_{c,i,t},\bar{u}_{d,i,t},\bar{d}_{f,i,t},\bar{\pi}_{i,t},\bar{g}_{i,t},\bar{c}_{i,t}$) denote the optimal solution of \textbf{P3} and \textbf{P4}, respectively. Since the adoption of ramping constraints in \textbf{P4} would or would not change the value of $\bar{c}_{i,t}$, three cases would be incurred.\\
\textbf{Case 1}: when $c_{i,t}^*=\bar{c}_{i,t}$, we have $y_{3,i}^{*}=\bar{y}_{4,i}$, where $y_{3,i}^{*}$ and $\bar{y}_{4,i}$ are the optimal objective value associated with the SMG $i$, respectively.\\
\textbf{Case 2}: when $c_{i,t}^*>\bar{c}_{i,t}$, the effective range of $c_{i,t}$ in \textbf{P4} is $\max\{c_{i,t-1}-\epsilon_i c_{i,\max},0\}\leq c_{i,t}\leq c_{i,t-1}+\epsilon_i c_{i,\max}$. We choose a feasible solution to \textbf{P4} as follows, i.e., ($x_{i,q,t}^*$, $u_{c,i,t}^*,u_{d,i,t}^*,d_{f,i,t}^*,\pi_{i,t}^*,g_{i,t}^*+c_{i,t}^*-c_{i,t-1}-\epsilon_i c_{i,\max},c_{i,t-1}+\epsilon_i c_{i,\max}$), which means that the conventional generator must generate less energy due to the ramping constraint and more energy should be purchased from the main grid $i$ to balance power. Then, we have $\bar{y}_{4,i}-y_{3,i}^{*}\leq V(1-\epsilon_i)c_{i,\max}X_{i,\max}$.\\
\textbf{Case 3}: when $c_{i,t}^*<\bar{c}_{i,t}$, the effective range of $c_{i,t}$ in \textbf{P4} is $c_{i,t-1}-\epsilon_i c_{i,\max}\leq c_{i,t}\leq \min\{c_{i,\max},c_{i,t-1}+\epsilon_i c_{i,\max}\}$. Set a feasible solution of \textbf{P4} as ($x_{i,q,t}^*$, $u_{c,i,t}^*,u_{d,i,t}^*,d_{f,i,t}^*,\pi_{i,t}^*,g_{i,t}^*+c_{i,t}^*-c_{i,t-1}+\epsilon_i c_{i,\max},c_{i,t-1}-\epsilon_i c_{i,\max}$), which means that the conventional generator must generate more energy due to the ramping constraint and more energy should be sold to the main grid $i$ to balance power. As a result, $\bar{y}_{4,i}-y_{3,i}^{*}\leq V(1-\epsilon_i)c_{i,\max}A_{i,\max}'$.\\
In summary, $\bar{y}_4\leq y_3^{*}+\sum\nolimits_{i=1}^N V(1-\epsilon_i)c_{i,\max}\gamma_{i,\max}$, which completes the proof.
\item Let ($x_{i,q,t}$, $u_{c,i,t},u_{d,i,t},d_{f,i,t},\pi_{i,t},g_{i,t},c_{i,t}$) and ($\hat{x}_{i,q,t}$, $\hat{u}_{c,i,t},\hat{u}_{d,i,t},\hat{d}_{f,i,t},\hat{\pi}_{i,t},\hat{g}_{i,t},\hat{c}_{i,t}$) denote the optimal solution of \textbf{P5} and the proposed algorithm, respectively. According to the online adjustment in Algorithm 1, we have $\Gamma_{p,t}-\Gamma_{5,t}\leq \sum\nolimits_{i=1}^N\big(\sigma_i(\hat{u}_{c,i,t}^2+\hat{u}_{d,i,t}^2)+\delta_{1,i}\hat{c}_{i,t}^2+\delta_{2,i}\hat{c}_{i,t}\big)\leq \Omega_2$, where
    $\Gamma_{5,t}$ and $\Gamma_{p,t}$ are the values of $\Gamma_t$ corresponding to the solutions of \textbf{P5} and the proposed algorithm, respectively.
\item Let $y_1$ and $y_2$ denote the optimal solution of \textbf{P1} and \textbf{P2}, respectively. Since \textbf{P2} is a relaxation of \textbf{P1}, we have $y_2\leq y_1$. Since \textbf{P5} is a relaxation of \textbf{P4}, we have
\begin{align} \label{af_7}
&\Delta_t + V\mathbb{E}\{\Gamma_{5,t}|\boldsymbol{\Theta_t}\} \\
\leq &\Delta_t + V\mathbb{E}\{\Gamma_{4,t}|\boldsymbol{\Theta_t}\} \\
\leq &\Omega_1+\Omega_0+V\mathbb{E}\{\tilde{\Gamma}_{3,t}|\boldsymbol{\Theta_t}\}\nonumber \\
&+\mathbb{E}\{\sum\limits_{i=1}^{N}\sum\limits_{q=1}^{M_i}wQ_{i,q,t}(\pi_{i,q,t}-x_{i,q,t}^*)|\boldsymbol{\Theta_t}\} \nonumber \\
&+\mathbb{E}\{\sum\limits_{i=1}^{N}\sum\limits_{q=1}^{M_i}wH_{i,q,t}(\varepsilon_{i,q}-x_{i,q,t}^*)|\boldsymbol{\Theta_t}\} \nonumber \\
&+\mathbb{E}\{\sum\limits_{i=1}^{N}Z_{i,t}(\eta_{c,i} u_{c,i,t}^*-\frac{1}{\eta_{d,i}}u_{d,i,t}^*)|\boldsymbol{\Theta_t}\} \\
\leq &\Omega_1+\Omega_0+Vy_2\\
\leq &\Omega_1+\Omega_0+Vy_1,
\end{align}
where $\Gamma_{4,t}$ and $\Gamma_{3,t}$ are the values of $\Gamma_t$ corresponding to the solutions of \textbf{P4} and \textbf{P3}, respectively; $x_{i,q,t}^*,u_{c,i,t}^*,u_{d,i,t}^*$ are the elements in the solution vector of \textbf{P3}; (58) is derived by the part 5 of Theorem 1; (59) is obtained by incorporating the results of a stationary, randomized control strategy associated with \textbf{P2}\cite{Neely2010}. By arranging the both sides of the above equations, we have $\mathbb{E}[\Delta_t] + V\mathbb{E}[\Gamma_{5,t}] \leq \Omega_1+\Omega_0+Vy_1$. Continually, we have $V {\sum\nolimits_{t = 0}^{T - 1} {\mathbb{E}\{\Gamma_{5,t}\} } } \leq \Omega_1T+YT+ VTy_1 - \mathbb{E}\{ L_T\}  + \mathbb{E}\{ L_0\}$. Dividing both side by $VT$, and taking a \text{lim sup} of both sides. Then, let $T \to \infty$, we have $\mathop {\lim \sup }\nolimits_{T \to \infty } \frac{1}{T}{\sum\nolimits_{t = 0}^{T - 1} {\mathbb{E}\{\Gamma_{5,t}\}}} \le y_1+\frac{\Omega_0+\Omega_1}{V}$. By taking the part 6 of Theorem 1 into consideration, we have $\mathop {\lim \sup }\nolimits_{T \to \infty } \frac{1}{T}{\sum\nolimits_{t = 0}^{T - 1} {\mathbb{E}\{\Gamma_{p,t}\}}} \le y_1+\Omega_2+\frac{\Omega_0+\Omega_1}{V}$, which completes the proof.
\end{enumerate}
\end{IEEEproof}

\end{document}